\documentclass[prb,twocolumn,amsmath,amssymb,aps,superscriptaddress,longbibliography]{revtex4-2}
\usepackage{graphicx}
\usepackage{dcolumn}
\usepackage{bm}
\usepackage{verbatim}
\renewcommand\Re{\operatorname{Re}}
\renewcommand\Im{\operatorname{Im}}

\begin{document}

\title{Contribution of Collective Excitations to Third Harmonic Generation in Multiband Superconductors: the case of MgB\textsubscript{2}}

\author{Jacopo Fiore}
\affiliation{Department of Physics and ISC-CNR, ``Sapienza'' University of Rome, P.le
A. Moro 5, 00185 Rome, Italy}
\author{Mattia Udina}
\affiliation{Department of Physics and ISC-CNR, ``Sapienza'' University of Rome, P.le
A. Moro 5, 00185 Rome, Italy}
\author{Marco Marciani}
\affiliation{Department of Physics and ISC-CNR, ``Sapienza'' University of Rome, P.le
A. Moro 5, 00185 Rome, Italy}
\author{Goetz Seibold}
\affiliation{Institut f\"ur Physik, BTU Cottbus-Senftenberg, PBox 101344, 03013 Cottbus,
Germany}
\author{Lara Benfatto}
\affiliation{Department of Physics and ISC-CNR, ``Sapienza'' University of Rome, P.le
A. Moro 5, 00185 Rome, Italy}

\date{\today}

\begin{abstract}
Multiband superconductors can host collective excitations with marked differences with respect to their single-band counterpart. We first study the spectrum of collective amplitude fluctuations in a clean two-bands superconductor, showing that the spectral weight of the Higgs mode rapidly deviates from the naive extension of the single band case as the interband coupling is turned on. These results are then used to critically analyze the non-linear optical response in MgB\textsubscript{2}, providing an explanation for the apparently contradictory results of recent experiments, pointing towards a selective relevance either of the Leggett mode or of the amplitude fluctuations at twice the lower gap. By using exact numerical simulations and realistic estimate of disorder we compute the relative contribution of the quasiparticle, amplitude and phase fluctuations to the non-linear optical response. We show that at low pumping frequency only the resonance at twice the smaller gap emerges, as due to the BCS response, while the Leggett mode dominates only in a narrow range of higher pumping frequencies matching its low-temperature value. Our findings provide a fresh perspective on the potential of non-linear THz spectroscopy to detect collective modes in other multiband systems, as e.g. iron-based superconductors.
\end{abstract}

\maketitle

\section{Introduction}

Multiband superconductivity has been reported for a variety of interesting systems, and it has been linked to both conventional (phonon-mediated) and unconventional pairing mechanisms. To the former category belongs the benchmark case of MgB\textsubscript{2}\cite{liu_prl2001}, that has been historically the first well-studied case of multiband superconductors, but also more recent examples as superconducting (SC) interfaces between insulating oxides\cite{biscaras_prl12,singh_natmat19} or layered dichalcogenides\cite{iwasa_natphys15}. To the latter category belong the widely studied families of iron-based superconductors, where pairing originates from electronic correlations, as possibly due to the exchange of spin fluctuations among quasi-nested hole-like and electron-like bands\cite{kotliar_review22,mazin_prl08,chubukov_rev12}. Even though rather often metals display multiple bands at the Fermi level, the characteristic feature of multiband superconductors that is common to all the above-mentioned cases is the presence of markedly different SC gaps $\Delta_\alpha$ on the various sheets of the Fermi surface, labelled here with a band index $\alpha$. This fact can lead to a series of spectroscopic signatures rather different from the single-band case, since the ratio $\Delta_\alpha/T_c$ between the gap values and the critical temperature $T_c$ can deviate from the standard, single-band BCS case, even in the presence of weak-coupling pairing\cite{dolgov_prb09,benfatto_prb2009}. This effect manifests in all the probes sensitive to thermal activation of single-particle 
excitations, as e.g. in the specific heat, and in the spectroscopic probes sensitive to the particle-hole excitations above the $2\Delta_\alpha$ optical gap in each band. In both cases, the observed behavior differ from the single-band case, and very often only the lower optical gap can be resolved in the optical absorption, since this is the only one in the dirty limit \cite{ummarino_julich,kuzmenko_ssc2001,quilty_prl2003,tu_prb10,charnukha_prb11}. 

In addition to the above modifications of the quasiparticle excitations, multiband superconductors host also a spectrum of collective fluctuations much richer than in the single-band case\cite{leggett,sharapov_epl02,aoki_prb11,babaev_prb11,marciani_prb2013,maiti_prb13,cea_prb2016,manske-leggett-natcomm16,murotani_2bands_prb17,giorgianni_natphys2019,shimano_prb19,haenel_prb2021}. Indeed, by considering for the sake of discussion the  two-bands case, each order parameter admits both amplitude and phase fluctuations below $T_c$. In the phase sector, besides the Goldstone mode of the broken $U(1)$ symmetry, associated with the fluctuations of the overall $\theta_1+\theta_2$ phase, a second so-called Leggett mode connected to relative phase fluctuations $\theta_1-\theta_2$ emerges\cite{leggett}. While the Goldstone mode is dynamically pushed to the plasma energy by its coupling to charge fluctuations, the mass $\omega_L$ of the Leggett mode is only controlled by the pairing strength\cite{leggett,sharapov_epl02}. For weak interband coupling it can occur below the lower optical gap, and it can even soften completely in systems with three bands at the verge of a time-reversal-symmetry-breaking transition\cite{aoki_prb11,babaev_prb11,marciani_prb2013,maiti_prb13}. For what concerns the amplitude fluctuations, the general behavior is rather different from the single-band case, where the spectral function of the so-called Higgs mode has a well-defined resonance at twice the gap value. As it has been already noticed in previous works\cite{manske-leggett-natcomm16,murotani_2bands_prb17,shimano_prb19,haenel_prb2021}, for multiband superconductors amplitude fluctuations in each band carry information on the optical gaps of both the bands. 

This rich phenomenology, which already challenges the interpretation of experiments with conventional spectroscopies, makes the understanding of high-field THz spectroscopies rather complex\cite{cea_prb2016,murotani_2bands_prb17,shimano_prb19,haenel_prb2021}. An example is provided by recent findings in MgB\textsubscript{2}, where both time-resolved protocols\cite{giorgianni_natphys2019} and third-harmonic generation in transmission\cite{kovalev_prb2021} have been used to investigate the non-linear optical response triggered by strong THz light pulses. As it has been widely discussed within the context of single-band models\cite{aoki_prb15,cea_prb16,aoki_prb16,silaev_prb2019,tsuji_prr20,seibold_prb2021} the main essence of these experiments is to trigger a non-linear current $j^{NL}$ which scales with the third power of the THz field $E$, $j^{NL}\sim \chi^{(3)} E^3$, where $\chi^{(3)}$ is the non-linear optical kernel and time convolution has been omitted for simplicity. Whenever the $\chi^{(3)}$ admits a marked maximum (resonance) at a given frequency $\omega_{res}$ one can observe, under specific experimental conditions, oscillations in the differential probe signal at $\omega_{res}$ as a function of the pump-probe time-delay $t_{pp}$ in time-resolved experiments, or enhanced THG for a multicycle light pulse centred at approximately $\omega_{res}/2$ \cite{udina_prb19,udina_faraday22}. In general, $\omega_{res}$ can be identified with any collective excitation of the system non-linearly coupled to the light. In a superconductor this is represented either by the threshold for the BCS particle-hole continuum at the optical gap $2\Delta$, or by the collective amplitude and phase modes. As a consequence, in a multiband superconductor one could expect in principle multiple signatures: quasiparticles or Higgs fluctuations at $2\Delta_\alpha$ of each band, and the Leggett mode at $\omega_L$. In MgB\textsubscript{2} $\omega_L$ lies in between the lower $2\Delta_{\pi}$ and the higher $2\Delta_\sigma$ optical gap, as it has been proven by spontaneous Raman measurements\cite{blumberg_prl07}. On the other hand, non-linear optics reported possible resonances in $\chi^{(3)}$ at either $\omega_L$\cite{giorgianni_natphys2019} or $2\Delta_\pi$\cite{kovalev_prb2021}, with the two features appearing mutually exclusive in the two experimental set-ups, challenging their interpretation. 

\begin{figure}
\centering
\includegraphics[width=0.75\columnwidth]{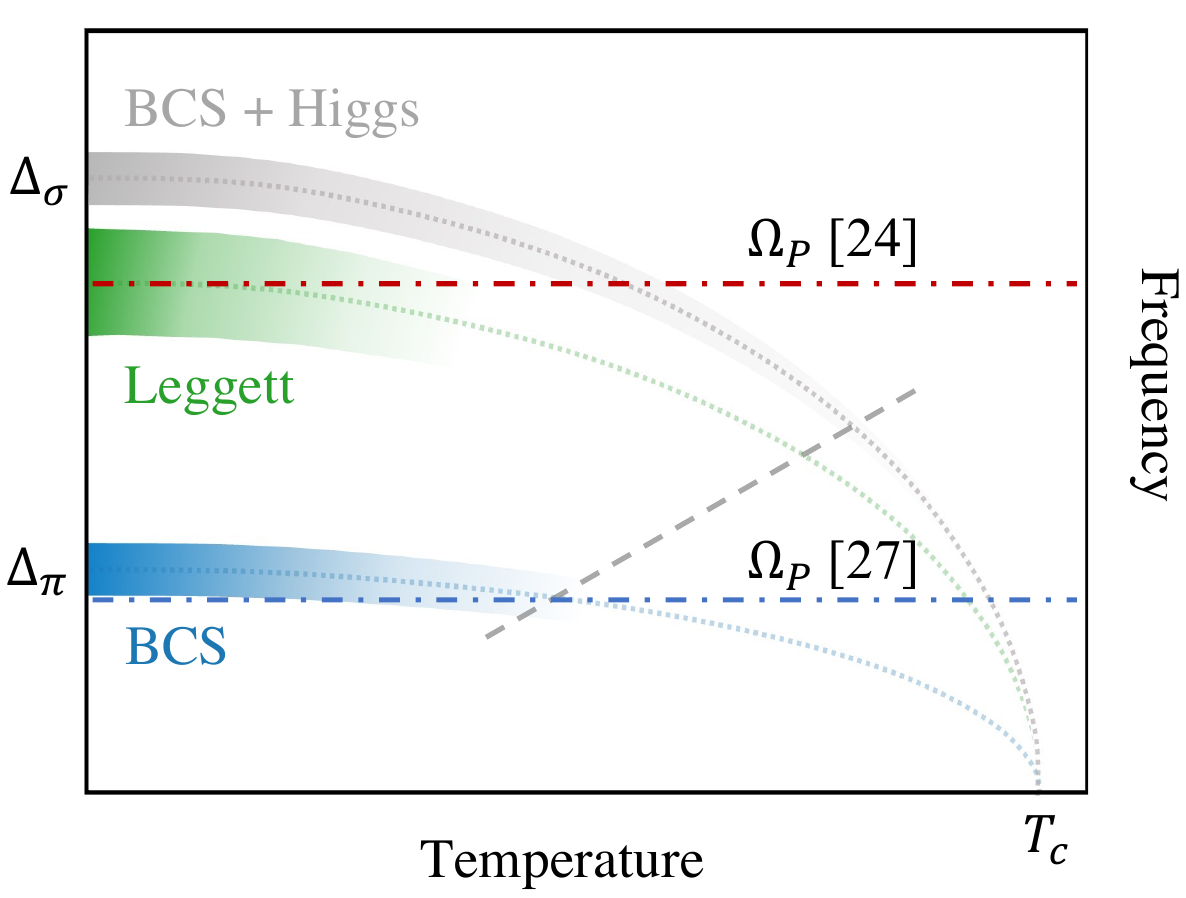}
\caption{Sketch of the various contributions to THG stemming from the spectrum of the collective fluctuations in MgB\textsubscript{2}. Dotted lines denote the temperature dependence of the two order parameters and of the Leggett mode, while shaded areas mark the typical frequency broadening of the corresponding resonances in the non-linear optical kernel. Around $\Delta_\sigma(T)$ both the BCS and Higgs fluctuations have a sizeable weight, while around $\Delta_\pi(T)$ BCS fluctuations dominate. The dashed line marks the temperatures $T^{*}_{\pi}$ and $T^{*}_{\sigma}$ such that $\Delta_{\pi}(T^{*}_{\pi})/T^{*}_{\pi}=\Delta_{\sigma}(T^{*}_{\sigma})/T^{*}_{\sigma}\approx2$, that is the condition at which the BCS/Higgs resonances are smeared out and damped by thermal fluctuations. The dashed-dotted lines denote the pumping frequencies of the experiment by Giorgianni et al. \cite{giorgianni_natphys2019} (red) and by Kovalev et al. \cite{kovalev_prb2021} (blue).}
\label{fig:fig0}
\end{figure}

In the present manuscript we provide a detailed analysis of the non-linear response of multiband superconductors, with the twofold aim from one side to clarify the nature of the collective fluctuations, in particular the Higgs mode, and from the other side to establish the experimental conditions for their detection via non-linear optics, that has been recently applied both to MgB\textsubscript{2}\cite{giorgianni_natphys2019,kovalev_prb2021} and iron-based superconductors\cite{wang_natcomm21,shimano_commphys21,grasset_npjqm22}. For what concerns the amplitude fluctuations, we will show that the redistribution of spectral weight between the two peaks at $2\Delta_\alpha$ in the spectral functions depends crucially on the structure of the coupling matrix. This leads in particular to a pile up of the spectral weight at twice the largest gap for MgB\textsubscript{2} and to a less straightforward redistribution for iron-based superconductors, crucially depending on the system parameters. In the latter case, an analogous effect has been observed for the Leggett mode, whose signature moves at twice the largest case as the SC coupling becomes predominantly interband\cite{cea_prb2016}. Once established the nature of the collective fluctuations of each order parameter, we will critically analyze their manifestation in the non-linear optical response for the specific case of MgB\textsubscript{2}. Indeed, as it has been widely discussed in the single-band case\cite{silaev_prb2019,tsuji_prr20,seibold_prb2021}, the relative contribution of both quasiparticle and Higgs-mode fluctuations to the $\chi^{(3)}$ response at $2\Delta_\alpha$ critically depend on the disorder level of the sample, with the former ones dominating at weak and intermediate disorder and the latter one dominating at very strong disorder. As a consequence, no definitive conclusion can be drawn without an accurate {\em quantitative} estimate of the disorder level of the sample, that can be obtained by a direct inspection of linear spectroscopy. We will use the experimentally estimated values $\gamma_\alpha/(2\Delta_\alpha)$ in each band to extract the contribution of each mode to the non-linear kernel, with $\gamma_\alpha$ being the scattering rate for carriers in the $\alpha$ band. In particular we extend the clean-limit calculations of Ref. \cite{cea_prb2016,giorgianni_natphys2019} to an effective model accounting for the disorder-mediated coupling of BCS, amplitude and phase fluctuations to light, in order to establish the dominant contributions to the non-linear response as a function of the temperature and the frequency of the light pulse, as summarized in Fig. \ref{fig:fig0}. We then show that the apparent discrepancy among the two experiments of Ref. \cite{giorgianni_natphys2019} and \cite{kovalev_prb2021} is purely linked to the different experimental conditions. In particular 
the strong signatures at $2\Delta_\pi$ observed in Ref. \cite{kovalev_prb2021}, where the pumping frequency is of the order of $\Delta_\pi(T\approx 0)$,  should be ascribed to BCS quasiparticle excitation in the $\pi$ band. In contrast, the measurements of Ref. \cite{giorgianni_natphys2019} are performed with a larger frequency, matching one half of the low-temperature value of the Leggett mode, making it the most relevant contribution to the non-linear response, see Fig. \ref{fig:fig0}. 
% We notice that this is also the reason why the Leggett-mode contribution is subdominant in the theoretical simulations of Ref. \cite{shimano_prb19}, that have been carried out in the same condition, i.e. $2\Omega_P\ll \omega_L(T\approx0)$. 
Finally, we show how multiband MgB\textsubscript{2} is the ideal system to estimate the effects of the linear-response screening in the SC state to correctly match the theoretical predictions with the experimental results. This issue has been recently raised by the analysis of experiments on $d$-wave cuprate superconductors\cite{kaiser_natcomm20}, but seems not relevant for conventional $s$-wave systems as NbN. Here we show that in the multiband case, due to the presence of a low $\Delta_\alpha/T_c$ ratio in one band, screening effects must be included to capture the correct temperature dependence of the THG, in analogy with what is seen in the case of a nodal single-band superconductor. 

The plan of the manuscript is the following. In Sec. \ref{sec:2} we study the spectrum of the amplitude fluctuations in a two-band superconductor, and we show representative results for the case of MgB\textsubscript{2} and iron-based superconductors. In Sec. \ref{sec:3} we introduce the formalism to study the non-linear optical response of MgB\textsubscript{2} in the clean or disordered case. in Sec. \ref{sec:4} we define a proper phenomenological model to account for disorder effect, where the couplings between light and electronic fluctuations are computed numerically via an exact procedure. In Sec. \ref{sec:5} we present the theoretical results and we compare them with the existing experimental data. Sec. \ref{sec:6} contains the final discussion and conclusions. The Appendices contain additional material. Appendix \ref{app:a} presents a detailed study of the experimental optical conductivity  of MgB\textsubscript{2}, aimed at extracting realistic estimates for the ratio between scattering rate and SC gap in each band. Appendix \ref{app:b} contains additional details on the normalization procedure for the light-matter coupling constants while Appendix \ref{app:c} explains the effect of short-range interactions on the value of the Leggett-mode frequency.

\section{Amplitude fluctuations: from the intraband to the interband dominated case.}
\label{sec:2}
As we mentioned in the introduction, the recent experimental advances in the use of strong THz pulses to probe SC systems has been responsible for an increasing theoretical interest on the nature of collective fluctuations of the SC order parameter. In particular the observation of an increased non-linear optical response below $T_c$ calls for a deeper understanding of the nature of collective excitations, and their coupling to the electric field. Indeed, as discussed in details e.g. in Ref. \cite{udina_prb19,seibold_prb2021}, in order to understand the experiments one has to address two separate problems: (i) the nature of the collective excitations, encoded in their spectral function, and (ii) their coupling to light. This is the scheme developed in detail to study more conventional collective excitations like phonons\cite{merlin_ssc97,maehrlein_prb18}: their contribution to the non-linear response depends on their nature (oscillation frequency and damping) and on their coupling to light, dictated by symmetry (with a distinction among infrared or Raman-active phonons). In the case of single-band superconductors the ongoing discussion on the contribution of the Higgs mode to THG has been focused on the second aspect: indeed, while it is well understood that the Higgs mode peaks at twice the SC gap in a single-band supercondutor, its non-linear coupling to light has been shown to depend on disorder, being negligible in the clean limit\cite{cea_prb16} and sizeable in the strong-disorder limit\cite{silaev_prb2019,tsuji_prr20,seibold_prb2021}. However, as we shall see in the present section, for a multiband superconductor also the nature of the Higgs fluctuations is not trivial, being strongly affected by the nature of the interband coupling. 
To address this issue, we introduce a two-bands ($\alpha,\beta=1,2$) BCS hamiltonian, in close analogy with previous works \cite{manske-leggett-natcomm16,cea_prb2016,murotani_2bands_prb17,giorgianni_natphys2019,shimano_prb19,haenel_prb2021},
\begin{equation}
\label{eqham}
H=\sum_{\alpha\mathbf{k}\sigma}\xi_{\alpha\mathbf{k}}c_{\alpha\mathbf{k}\sigma}^{\dagger}c_{\alpha\mathbf{k}\sigma}-\sum_{\alpha\beta\mathbf{q}}g_{\alpha\beta}\Psi_{\alpha\mathbf{q}}^{\dagger}\Psi_{\beta\mathbf{q}},
\end{equation}
where $\Psi_{\alpha\mathbf{q}}^{\dagger}$ is written explicitly as
\begin{equation}
\Psi_{\alpha\mathbf{q}}^{\dagger}=\sum_{\mathbf{k}}c_{\alpha\mathbf{k}+\mathbf{q}/2\,\uparrow}^{\dagger}\,c_{\alpha-\mathbf{k}+\mathbf{q}/2\,\downarrow}^{\dagger}.
\end{equation}
The kinetic term contains $\xi_{\alpha\mathbf{k}}=\epsilon_{\alpha\mathbf{k}}-\mu$, where $\epsilon_{\alpha\mathbf{k}}$ is the band dispersion and $\mu$ is the chemical potential. We consider a continuum model with free-electron like dispersions of the form 
\begin{equation}
\epsilon_{\alpha\mathbf{k}}=s_{\alpha}\Bigl(\frac{k^2}{2m_{\alpha}}-\epsilon_{\alpha0}\Bigr),
\end{equation}
where $\epsilon_{\alpha0}$ is the distance of the bottom ($s_{\alpha}=1$, electron-like band) or top ($s_{\alpha}=-1$, hole-like band) of the band from the chemical potential in the normal state and $m_{\alpha}$ is the band mass. Such an approximation is good both for MgB\textsubscript{2} \cite{kortus_prl2001} and pnictides \cite{mazin_pc2009}. The interacting term contains the symmetric matrix $g_{\alpha\beta}$ that includes an effective $s$-wave interaction among Cooper pairs belonging to the same band (diagonal entries) or different band (off-diagonal entries). When the determinant $\lvert g\rvert$ of the interaction matrix is positive (negative) the pairing is said to be intraband (interband) dominated. 

To study the collective fluctuations we derive the effective action for the bosonic pairing fields, by following a standard procedure outlined in several manuscripts, see e.g. Ref. \cite{sharapov_epl02,marciani_prb2013,cea_prb2016,cea_prb18} for the multiband case. We perform a canonical Hubbard-Stratonovich (HS) transformation to decouple the interaction along the superconducting channel of each band by introducing the complex fields $\Delta_{\alpha}$. The resulting action is then quadratic in the fermionic variables that can be integrated out, obtaining the effective action for the HS fields. We separate then the mean-field contribution from the fluctuating part truncated at Gaussian order in the HS fields. Minimization of the mean-field contribution leads to the equations for the gap amplitudes that read
\begin{equation}
\label{eq:mfgap}
(\Pi-g^{-1})\Delta=0,
\end{equation}
where $\Delta=(\Delta_1,\Delta_2)$ can be chosen real due to the $U(1)$ symmetry of the action and being $\Pi-g^{-1}$ a symmetric matrix. The term $\Pi_{\alpha\beta}=\Pi_{\alpha}\delta_{\alpha\beta}$ is the so-called Cooper's bubble reading in terms of $E_{\alpha\mathbf{k}}=\sqrt{\xi_{\alpha\mathbf{k}}^2+\Delta_{\alpha}^2}$
\begin{equation}
\Pi_{\alpha}=\sum_{\mathbf{k}}\frac{\tanh{(\beta E_{\alpha\mathbf{k}}/2)}}{2E_{\alpha\mathbf{k}}}.
\end{equation}
The remaining action for the uniform ($\lvert\mathbf{q}\rvert=0$) gaussian fluctuations of the mean-field gaps can be written as 
\begin{equation}
\label{eq:genflu}
S_{\text{fluc}}\sim\Re{\delta \Delta}^TM_A\Re{\delta \Delta}+\Im{\delta\Delta}^TM_P\Im{\delta\Delta}.
\end{equation}
The first (second) contribution involves fluctuations $\delta \Delta$ of the real (imaginary) part of the order parameter with respect to the mean-field gaps. Since we chose a real mean-field gap, the first term corresponds to amplitude fluctuations, whose propagator is given by $M_A^{-1}$. Being then $\Im\delta{\Delta}=(\Delta_1\delta\theta_1,\Delta_2\delta\theta_2)$, the second contribution describes the fluctuations of the phase $\theta_\alpha$ in each band, governed by the propagator $M_P^{-1}$. We notice that in this continuum model the amplitude and phase fluctuations are decoupled at $\lvert\mathbf{q}\rvert=0$\cite{cea_prb2016}. 

We then study the behavior of the amplitude fluctuations of a generic two-bands superconductor as a function of the interband pairing $g_{12}$. At gaussian level these are expressed as
\begin{equation}
\label{eq:aflu}
\langle\lvert\delta\Delta_{\alpha}\rvert\lvert\delta\Delta_{\beta}\rvert\rangle=(M_A^{-1})_{\alpha\beta},
\end{equation}
where the matrix $M_A$ reads,
\begin{equation}
M_A=
\begin{pmatrix}
\chi_{1}^{\sigma_1\sigma_1}+2g_{11}^{-1} & 2g_{12}^{-1} \\
2g_{12}^{-1} & \chi_{2}^{\sigma_1\sigma_1}+2g_{22}^{-1} \\
\end{pmatrix}.
\end{equation}
The fermionic bubble $\chi_{\alpha}^{\sigma_1\sigma_1}$ reads explicitly, introducing the fermionic Matsubara frequencies $i\omega_n$,
\begin{equation}
\chi_{\alpha}^{\sigma_1\sigma_1}(i\omega_n)=(4\Delta_{\alpha}^2-(i\omega_n)^2)F_{\alpha}(i\omega_n)-2\Pi_{\alpha},
\end{equation}
where the function $F_{\alpha}(i\omega_n)$ is defined as
\begin{equation}
\label{eq:falpha}
F_{\alpha}(i\omega_n)=\sum_{\mathbf{k}}\frac{\tanh{(\beta E_{\alpha\mathbf{k}}/2)}}{E_{\alpha\mathbf{k}}(4E_{\alpha\mathbf{k}}^2-(i\omega_n)^2)}.
\end{equation}

To put in evidence the main difference between the single and the two-bands cases, let us write the determinant of the matrix $M_A$
\begin{widetext}
\begin{equation}
\label{eq:det}
\lvert M_A\rvert=(4\Delta_1^2-\omega^2)F_1(\omega)(4\Delta_2^2-\omega^2)F_2(\omega)+\frac{2g_{12}}{\lvert g\rvert}\frac{\Delta_2}{\Delta_1}(4\Delta_2^2-\omega^2)F_2(\omega)+\frac{2g_{12}}{\lvert g\rvert}\frac{\Delta_1}{\Delta_2}(4\Delta_1^2-\omega^2)F_1(\omega).
\end{equation} 
\end{widetext}
A zero of the determinant signals the presence of a well-defined mode of the system. Since $(4\Delta_{\alpha}^2-\omega^2)F_{\alpha}(\omega)=0$ at $\omega=2\Delta_{\alpha}$, it is possible to get a zero in (\ref{eq:det}) if either the two bands are identical or the interband coupling is zero, otherwise the determinant is always nonzero and we only find a finite minimum at twice the amplitude of the two gaps. 

%Let us make this point more clear with respect to the single band case. In that situation it has been already pointed out \cite{cea_prl2015} that the propagator for the amplitude fluctuations has nothing to do with a relativistic-type resonance
%\begin{equation}
%\langle\lvert\Delta\rvert^2\rangle\sim\frac{1}{4\Delta^2-\omega^2},
%\end{equation}
%that can be for instance obtained in a $U(1)$ Lorentz invariant generalization of the Ginzburg-Landau (GL) model. The correct expression from the microscopic theory presents a finite imaginary part for $\omega>2\Delta$, signaling the proliferation of quasiparticle-quasihole pairs that ultimately damp the resonance above twice the gap amplitude. This phenomenon occurs in the same way in a two-bands GL model. In addition, at finite interband coupling the same GL model would predict two \textit{zeros} (not two \textit{minima}) in the determinant of $M_A$, introducing two spurious modes in the system that are indeed not foreseen from the microscopic theory. Conversely, the two-bands GL model allows to recover the correct spectrum of the Leggett mode in the phase sector.

In the latter case the spectra of the amplitude fluctuations in (\ref{eq:aflu}) presents two maxima and not two divergences. This is a first striking difference with respect to the single-band case, where the Higgs-mode fluctuations emerge out of a proper divergence of the spectral function at $2\Delta$, at least at weak and intermediate coupling \cite{cea_prl2015,cea_prb16}. Most importantly, at finite interband coupling both $\langle\lvert\delta\Delta_1\rvert^2\rangle$ and $\langle\lvert\delta\Delta_2\rvert^2\rangle$ present a finite spectral weight at both $\omega=2\Delta_1$ and $\omega=2\Delta_2$. To assess their relative weight as a function of interband coupling, we introduce the parameter $\lvert\eta\rvert<1$ \cite{cea_prb2016} such that 
\begin{equation}
\eta=\frac{g_{12}-\sqrt{g_{11}g_{22}}}{g_{12}+\sqrt{g_{11}g_{22}}}.
\end{equation}
For $-1<\eta<0$ ($0<\eta<1$) we are in the intraband (interband) dominated regime. Notice that $\eta=-1$ corresponds to a purely intraband interaction, i.e. to a diagonal coupling matrix. Instead for $\eta=1$ the product of the diagonal terms is zero. Together with $\eta$, we fix the ratio $\omega_c/\Delta_1$, where $\omega_c$ is the usual bosonic energy scale of the pairing, the ratio $\Delta_2/\Delta_1$ of the two gap amplitudes, and the ratio of the DOS at Fermi surface $N_2/N_1$. The entries of the coupling matrix will then be consistently determined in order to satisfy the gap equations (\ref{eq:mfgap}). 

\begin{figure}
\centering
\includegraphics[width=\columnwidth]{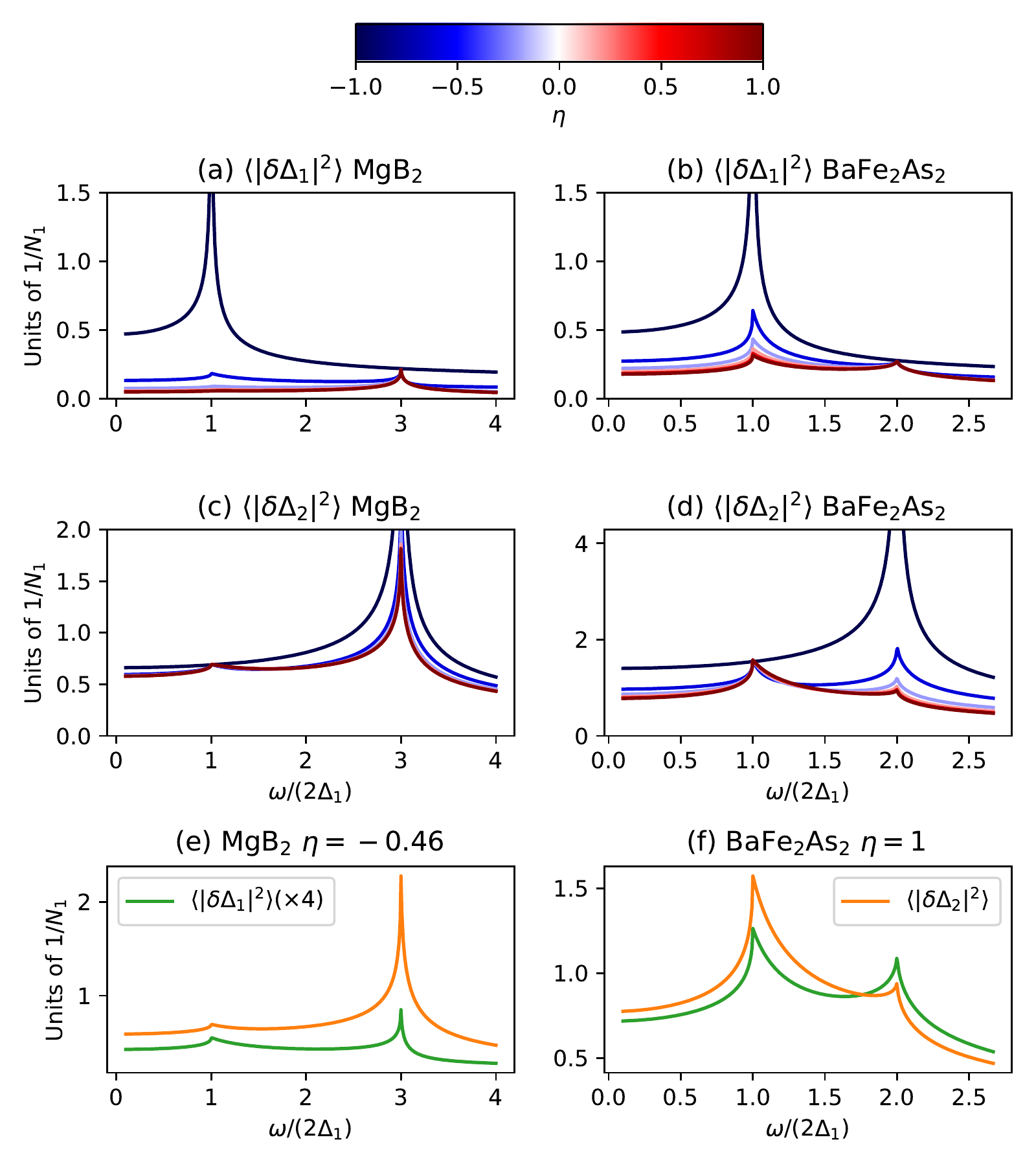}
\caption{The modulus of the amplitude fluctuations for the smaller (a) (c) and bigger (b) (d) gaps at fixed gap and DOS ratio $\Delta_2/\Delta_1=3$, $N_2/N_1=0.75$ (as appropriate for MgB\textsubscript{2}), (a) (b), and $\Delta_2/\Delta_1=2$, $N_2/N_1=0.35$ (as appropriate for BaFe\textsubscript{2}As\textsubscript{2}), (c) (d), as a function of interband coupling (blue tones for intraband and red tones for interband). Higgs fluctuations in MgB\textsubscript{2} (e) and in 122 pnictides (f) for realistic values of the interband coupling. Values of $\langle\lvert\delta\Delta_1\rvert^2\rangle$ are magnified $\times 4$ for clarity.}
\label{fig:fig1}
\end{figure}

We analyzed two cases in order to get an insight into the spectra of the amplitude fluctuations in MgB\textsubscript{2}, see left panels of Fig. \ref{fig:fig1}, and in a hole doped compound of the parent pnictide BaFe\textsubscript{2}As\textsubscript{2} (BKFA), see right panels of Fig. \ref{fig:fig1}. In the first case we use $N_2/N_1=0.75$ and $\Delta_2/\Delta_1=3$ \cite{golubov_jpcm2002, liu_prl2001}, where label $1$ refers to the $\pi$ band and $2$ to the $\sigma$ one. In the second case we set $N_2/N_1=0.35$ and $\Delta_2/\Delta_1=2$ \cite{benfatto_prb2009, mazin_pc2009}, with label $1$ assigned to the $\beta$ band (hole like, at $\Gamma$) and $2$ to the $\gamma$ one (electron like, at M).

We plot the values of $\langle\lvert\delta\Delta_1\rvert^2\rangle$ and $\langle\lvert\delta\Delta_2\rvert^2\rangle$ for different values of $\eta$. As we will see, in our phenomenological model for the non-linear response in the disordered case, the $\langle\lvert\delta\Delta_{\alpha}\rvert^2\rangle$ fluctuations associated with the most disordered band will be the most coupled to light and thus will dominate the Higgs sector. Therefore we avoid to report explicitly the behavior of $\langle\lvert\delta\Delta_1\rvert\lvert\delta\Delta_2\rvert\rangle$.

For both sets of band parameters, at $\eta=-1$ we have two well-defined divergences at $\omega=2\Delta_\alpha$ for $\langle\lvert\delta\Delta_\alpha\rvert^2\rangle$. We notice that as soon as the interband coupling is activated, the spectral weight of $\langle\lvert\delta\Delta_1\rvert^2\rangle$ at $\omega=2\Delta_2$ stays constant and the one at $\omega=2\Delta_1$ gradually decreases, while the complementary behavior is observed for $\langle\lvert\delta\Delta_2\rvert^2\rangle$. The situation for $\eta=1$ is different in the two cases of MgB\textsubscript{2} and pnictides. In MgB\textsubscript{2}, the spectral weight at $\omega=2\Delta_2$ is largely predominant over the one at $\omega=2\Delta_1$ for both channels, while in BKFA there is a comparable weight at both $\omega=2\Delta_1$ and $\omega=2\Delta_2$.

\begin{figure}
\centering
\includegraphics[width=\columnwidth]{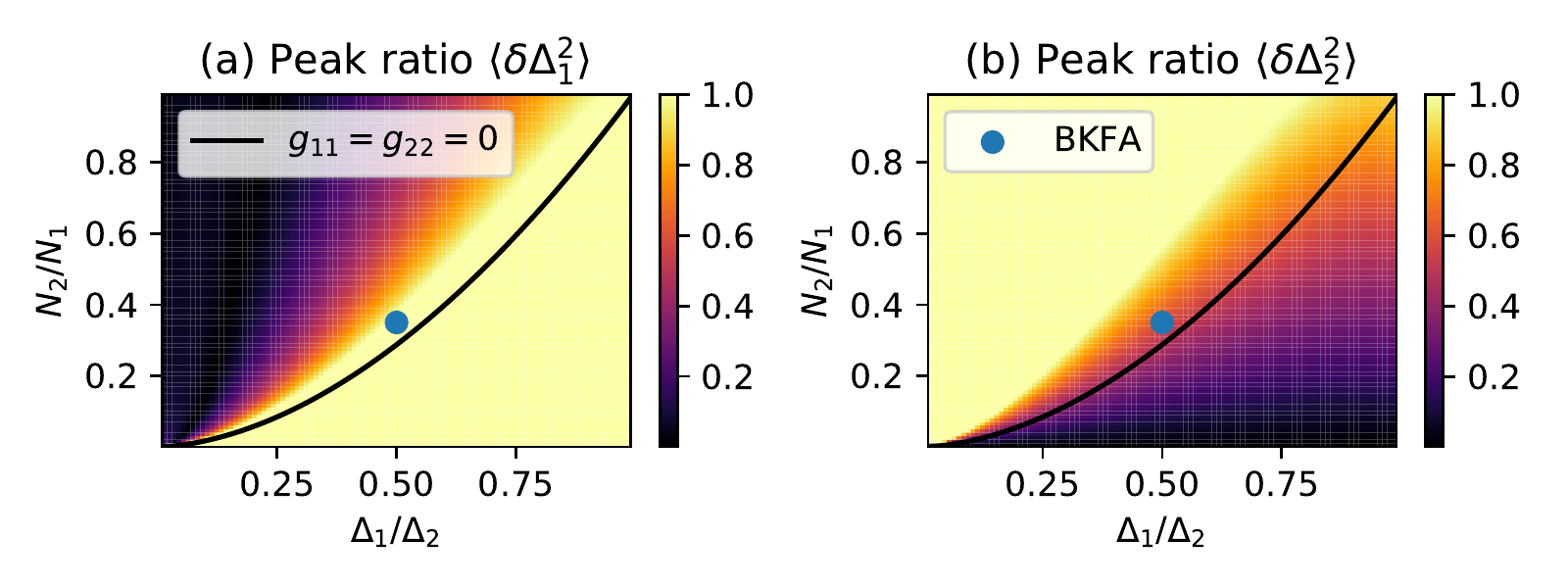}
\caption{The values $r_1$, equation (\ref{eq:r1}), in panel (a) and $r_2$, equation (\ref{eq:r2}), in panel (b) as a function of DOS anisotropy and gap ratio for dominated interband scattering $\eta=1$. Values greater than one have been saturated in order to better appreciate the region of anomalous behavior. Solid black line is the solution of gap equations for $g_{11}=g_{22}=0$. Blue dots are the position of the appropriate gap and DOS ratio for BKFA.}
\label{fig:fig2}
\end{figure}

To further clarify this behavior, we plot the ratio of the peaks at $\omega=2\Delta_1$ and $\omega=2\Delta_2$ for both channels for the extreme case $\eta=1$, as a function of the DOS and gap anisotropy. In panel (a) of Fig. \ref{fig:fig2} we plot
\begin{equation}
\label{eq:r1}
r_1=\frac{\langle\lvert\delta\Delta_1\rvert^2\rangle(\omega=2\Delta_1)}{\langle\lvert\delta\Delta_1\rvert^2\rangle(\omega=2\Delta_2)}, 
\end{equation}
while in panel (b) we show
\begin{equation}
\label{eq:r2}
r_2=\frac{\langle\lvert\delta\Delta_2\rvert^2\rangle(\omega=2\Delta_2)}{\langle\lvert\delta\Delta_2\rvert^2\rangle(\omega=2\Delta_1)}.
\end{equation}
We start by observing that in the case $\eta=1$ we do not necessarily have a completely off diagonal ($g_{11}=g_{22}=0$) coupling matrix. As can be seen by explicitly solving the gap equations, this happens only when the gap and DOS ratios satisfy
\begin{equation}
\frac{N_2}{N_1}=\Bigl(\frac{\Delta_1}{\Delta_2}\Bigr)^2\frac{\Pi_1}{\Pi_2}\approx\Bigl(\frac{\Delta_1}{\Delta_2}\Bigr)^2,
\end{equation}
where the approximation is valid in the limit $\omega_c\gg\Delta_1,\Delta_2$. The solution of this equation is plot as a solid black line in Fig. \ref{fig:fig2}. Changing the DOS ratio away from this condition at fixed gap ratio, or the converse, implies to increase one among $g_{11}$ or $g_{22}$ in order to keep the gap equation satisfied, while the other remains zero. Thus, one order parameter can now sustain amplitude fluctuations due to both interband and intraband scattering channels, while the other only through interband scattering, resulting in a predominant Higgs weight at twice the order parameter that has more scattering channels available. This is reflected in the peak ratios $r_1$ and $r_2$, that are respectively bigger than one roughly below ($g_{11}\neq0$, $g_{22}=0$) and above ($g_{11}=0$, $g_{22}\neq0$) the black line. 

In summary, the naive expectation that in a multiband superconductor the amplitude fluctuations in each band $\alpha$ have a sharp maximum at the optical gap $2\Delta_\alpha$ for the same band is in general not correct. In general, the absence of a zero in the determinant (\ref{eq:det}) makes the maxima of the spectral function for amplitude fluctuations in a multiband system much smoother than in a single-band case. In addition, signatures at one or the other optical gap can be strongly suppressed by means of the interband pairing. 

In panels (e) and (f) of Fig. \ref{fig:fig1} we show the spectral function of MgB\textsubscript{2} and 122 compounds, computed for realistic values of $\eta$. In the MgB\textsubscript{2} case, which is of interest in the following sections, we derive a suitable coupling matrix in order to reproduce the behavior in temperature of the $\pi$ gap of the sample used by Kovalev et al. \cite{kovalev_prb2021}. We fix the gap amplitude at zero temperature at $2\Delta_{\pi}^0=1.1$ THz and the critical temperature at $T_c=36$ K as measured by the experiment. The Debye frequency is set at $\omega_c=55$ meV, i.e. the average phonon frequency from ab initio calculations, and we take the calculated ratio between density of states per spin at the Fermi level, $N_{\sigma}/N_{\pi}=0.73$. We fix as well the level of interband coupling \cite{cea_prb2016} $\eta=-0.46$ desumed from the same ab initio calculations \cite{liu_prl2001}. We can solve the gap equations at $T=0$ and $T=T_c$ to obtain the adimensional coupling matrix elements $\tilde{g}_{\alpha\beta}=g_{\alpha\beta}N_{\beta}$, namely $\tilde{g}_{\pi\pi}=0.13$, $\tilde{g}_{\sigma\sigma}=0.29$, $\tilde{g}_{\pi\sigma}=0.06$ and $\tilde{g}_{\sigma\pi}=0.08$, a similar procedure being used in Ref. \cite{jin_prl2003}. %The obtained theoretical $\Delta_{\pi}(T)$ and $\Delta_{\sigma}(T)$ are shown in Fig. \ref{fig:gap}, where they are compared with the experimental $\pi$ gap behavior. 
The obtained values are comparable with the usual coupling matrices used for MgB\textsubscript{2} \cite{golubov_jpcm2002, liu_prl2001} when the renormalization due to the Coulomb pseudopotential is taken into account in the static limit \cite{ummarino_julich, ummarino_pc2004}, reading in the two-bands case $\tilde{g}_{\alpha\beta}\to({\tilde{g}_{\alpha\beta}-\mu_{\alpha\beta})/(1+\sum_{\beta}\tilde{g}_{\alpha\beta}})$.

As one can see in Fig. \ref{fig:fig1}, while for pnictides the amplitude fluctuations in each band retain a reasonable amount of spectral weight on both optical gaps, see panel (f), for MgB\textsubscript{2} the Higgs fluctuations in both bands have most of the spectral weight at the largest optical gap, i.e. at $2\Delta_\sigma$, see panel (e). This result must be taken in mind while discussing the Higgs-mode contribution in THz experiments, as we will do in the next section.

\section{Non-linear optical response in MgB\textsubscript{2}}
\label{sec:3}

Once clarified the nature of the amplitude fluctuations in a multiband superconductor, we want now to provide an estimate of their contribution to the non-linear optical response, and their relative importance as compared to BCS particle-hole excitations and to the Leggett mode, whose spectrum has been studied previously by several authors\cite{blumberg_prl07,sharapov_epl02,cea_prb2016}. We will focus on the case of MgB\textsubscript{2}, whose THz response at strong fields has been recently measured\cite{giorgianni_natphys2019,kovalev_prb2021}, by providing a phenomenological estimate of the contributions of the different superconducting fluctuations induced by the paramagnetic coupling of light triggered by disorder. 

Let us first recall the main idea behind the computation of the non-linear optical response, as it appears both in pump-probe experiments and in measurements of THG in transmission. As mentioned in the introduction and detailed in several manuscripts, we need to compute the non-linear current $j^{NL}\sim \chi^{(3)} E^3$. By using the effective action approach mentioned before, one can compute \cite{cea_prb16,cea_prb18,udina_prb19,haenel_prb2021} $j^{NL}\sim\delta S^{(4)}/\delta A\sim AKA^2$ as derivative of the fourth-order action $S^{(4)}[A]\sim A_i A_j K_{ij;kl}A_k A_l$ expressed in terms of the gauge field $\mathbf{A}$, that is minimally coupled to electrons in the original Hamiltonian (\ref{eqham}). After integrating out both the fermions and the collective modes one obtains the various contributions to the non-linear optical kernel $K$. The bare processes involve direct excitations of particle-hole processes by light, and will be denoted as BCS response in what follows, while vertex corrections represent the contribution of the collective modes, like Higgs and Leggett fluctuations. Finally, the intensity of the THG scales as $I_{THG}=K(2\Omega_p)$, where $\Omega_P$ is the central frequency of the THz light pulse, and the factor of two denotes the sum-frequency process behind non-linear driving of the system. We refer for further details to previous work \cite{cea_prb16,cea_prb2016,udina_prb19,seibold_prb2021}. 

\subsection{Clean system}

For the clean case the only BCS contribution is due to diamagnetic processes\cite{cea_prb16}, originating from a term like $\sim A^2 s_{\alpha}e^2/(2m_{\alpha})c^{\dagger}_{\alpha\mathbf{k}\sigma}c_{\alpha\mathbf{k}\sigma}$ in the microscopic Hamiltonian, corresponding to the usual diamagnetic contribution to electronic current. $K^{BCS}_{dia}$ in the two-bands model is simply given by the sum of two single band-like expressions, reading
\begin{equation}
\label{eq:bcs}
K^{BCS}_{dia}=\sum_{\alpha}\Bigl(\frac{s_{\alpha}e^2}{2m_{\alpha}}\Bigr)^2\chi_{\alpha}^{\sigma_3\sigma_3},
\end{equation}
where $\chi_{\alpha}^{\sigma_3\sigma_3}=4\Delta_{\alpha}^2F_{\alpha}(\omega)$. The corresponding bubble is shown in panel (a) of Fig. \ref{fig:fig3}. The enhancement of THG response when $2\omega=2\Delta_{\alpha}$, that we interpret as particle-hole excitations of BCS quasiparticles, is determined by the divergence of the $\chi_{\alpha}^{\sigma_3\sigma_3}$ bubble, the prefactor in round brackets in Eq. ({\ref{eq:bcs}) being only a measure of the strength of diamagnetic coupling with the two bands, since it includes the mediating charge $e$ and the inverse band mass $1/m_{\alpha}$. 

The Higgs contribution $K^{H}_{dia}$ in the clean two-bands case is given by a more involved expression reading 
\begin{equation}
\label{eq:diahiggs}
K^{H}_{dia}=\sum_{\alpha\beta}\Bigl(\frac{s_{\alpha}e^2}{2m_{\alpha}}\Bigr)\Bigl(\frac{s_{\beta}e^2}{2m_{\beta}}\Bigr)\chi_{\alpha}^{\sigma_3\sigma_1}\chi_{\beta}^{\sigma_3\sigma_1}(M_A^{-1})_{\alpha\beta}.
\end{equation}
We notice that if the interband coupling is zero we recover also here the sum of two single band-like amplitude fluctuations, as shown e.g. in Ref. \cite{cea_prb18}. The fermionic bubble $\chi_{\alpha}^{\sigma_3\sigma_1}$ reads
\begin{equation}
\label{eq:chi31}
\chi_{\alpha}^{\sigma_3\sigma_1}=2\Delta_{\alpha}\sum_{\mathbf{k}}\frac{\xi_{\alpha\mathbf{k}}\tanh{(\beta E_{\alpha\mathbf{k}}/2)}}{E_{\alpha\mathbf{k}}(4E_{\alpha\mathbf{k}}^2-(\omega+i0^{+})^2)}.
\end{equation}
As it has been discussed in Ref. \cite{cea_prb16}, such a bubble vanishes in the particle-hole symmetric cases since the integral (\ref{eq:chi31}) scales as $N_{\alpha}\int_{-\infty}^\infty d\xi G_{\alpha}(\xi)$, with $G_{\alpha}(\xi)$ odd function of $\xi$ and $N_{\alpha}$ DOS at the Fermi level of band $\alpha$. The coupling becomes however finite if one can account for a finite particle-hole asymmetry of the band, by retaining a linear term in the expression for the DOS, namely $N_{\alpha}(\xi)\approx N_{\alpha}+\xi/(2\epsilon_{0\alpha})$, so that
\begin{equation}
\label{eq:sig31}
\chi_{\alpha}^{\sigma_3\sigma_1}=\frac{\Delta_{\alpha}}{4\epsilon_{0\alpha}}\Bigl(2g_{\alpha\alpha}^{-1}+2g_{12}^{-1}\frac{\Delta_{\bar{\alpha}}}{\Delta_{\alpha}}-(4\Delta_{\alpha}^2-\omega^2)F_{\alpha}(\omega)\Bigr),
\end{equation}
with $\bar{\alpha}=2,1$ if $\alpha=1,2$. Even including such an asymmetry, for MgB\textsubscript{2} the $\chi^{\sigma_3\sigma_1}$ remains extremely small, since usually $\Delta_{\alpha}/\epsilon_{0\alpha}\ll1$,. The situation could be different for pnictides, where a similar prefactor $\Delta_{\alpha}/\epsilon_{0\alpha}$ is indeed obtained by considering a constant DOS but accounting for the fact that the bottom/top of the electron/hole like band lies inside the bosonic frequency range $\sim\omega_c$. In that case the integration runs over an asymmetric shell above and below the Fermi surface and Eq. (\ref{eq:sig31}) is finite. However, whenever the $\chi^{\sigma_3\sigma_1}$ bubble is not zero one should also retain the coupling of Higgs to the density/phase sector, that again suppresses this effect \cite{cea_prb16,cea_prb18}. As we shall see below, in what follow we will consider as main coupling of the Higgs to light the one induced by paramagnetic processes, so the structure of Eq. (\ref{eq:sig31}) will only be used to rescale such a coupling as a function of temperature. 

%We notice that expression (\ref{eq:sig31}) contains no divergences and has a maximum for $\omega=2\Delta_{\alpha}$. This means that the structure in frequency of the Higgs response is completely determined by $(M_A^{-1})_{\alpha\beta}(\omega)$, the $\chi_{\alpha}^{\sigma_3\sigma_1}(\omega)$ being only a measure of the strength of the coupling. 

While the Higgs-mode contribution is negligible in the clean limit, the coupling to phase fluctuations is crucial in order to guarantee gauge-invariance of the theory, especially for a system with an approximated parabolic band structure as the one considered here\cite{cea_prb16,cea_prb2016,maiti_prb17}. For the clean multiband case this has been done in Ref. \cite{cea_prb2016,maiti_prb17}, leading to a contribution to the non-linear kernel completely analogous to the Higgs case (\ref{eq:diahiggs}), see panel (b) of Fig. \ref{fig:fig3}, where now we should consider coupling with the phase fluctuations $\chi_{\alpha}^{\sigma_3\sigma_1}\to\chi_{\alpha}^{\sigma_3\sigma_2}$ and $(M_A^{-1})_{\alpha\beta}\to(M_P^{-1})_{\alpha\beta}$, with
\begin{equation}
M_P(i\omega_n)=
\begin{pmatrix}
\chi_{1}^{\sigma_2\sigma_2}(i\omega_n)+2g_{11}^{-1} & 2g_{12}^{-1} \\
2g_{12}^{-1} & \chi_{2}^{\sigma_2\sigma_2}(i\omega_n)+2g_{22}^{-1} \\
\end{pmatrix}.
\end{equation}
The resulting expression can be rearranged as
\begin{equation}
\label{phase}
\tilde{K}_{dia}^{L}=-K_{dia}^{BCS}+K_{dia}^{L}.
\end{equation}
The first term cancels the BCS fluctuations, so that in the clean case one is left only with the Leggett-mode contribution $K_{dia}^{L}$, which reads
\begin{equation}
\label{eq:kleg}
K_{dia}^{L}=\Bigl(\frac{s_{\sigma}e^2}{2m_{\sigma}}-\frac{s_{\pi}e^2}{2m_{\pi}}\Bigr)^2\frac{\lambda}{L},
\end{equation}
where $\lambda=8g_{\sigma\pi}^{-1}\Delta_{\sigma}\Delta_{\pi}$. The zeros of the denominator
\begin{equation}
\label{eq:legmode}
L=\omega^2-\lambda\frac{\chi_1^{\sigma_3\sigma_3}+\chi_2^{\sigma_3\sigma_3}}{\chi_1^{\sigma_3\sigma_3}\chi_2^{\sigma_3\sigma_3}},
\end{equation}
determine the dispersion of the Leggett mode. Notice that the Leggett-mode response only survives when the bands have different masses and/or opposite (hole-like vs electron-like) character, so that the term in round bracket of Eq. (\ref{eq:kleg}) is not zero. In contrast, when the two bands have same mass and same character the non-linear optical kernel is proportional to the density response of the system $K\sim \langle \rho \rho\rangle$, that must vanish at $\lvert\mathbf{q}\rvert=0$ and finite $\omega$ because of charge conservation\cite{cea_prb2016,maiti_prb17}. Since the BCS approximation by itself is not gauge invariant this result can only be achieved by integrating out the phase mode, as encoded indeed in the above Eq. (\ref{phase}). In summary, in the clean case where only diamagnetic-like coupling to light matters, only the Leggett mode contributes to the non-linear response, in agreement also with the experimental observation in spontaneous Raman experiments in the symmetric channel\cite{blumberg_prl07}, whose response function reduces in the effective-mass approximation to the same density-like correlation function considered here.

\subsection{Disordered system}

In presence of disorder the non-linear kernel acquires a finite contribution also from processes mediated by paramagnetic-like matter-light coupling terms, originating from terms like $\sim A ev_{\alpha\mathbf{k}}c^{\dagger}_{\alpha\mathbf{k}\sigma}c_{\alpha\mathbf{k}\sigma}$ in the microscopic Hamiltonian, with $v_{\alpha\mathbf{k}}$ band velocity, as shown by the diagrams (c) and (d) in Fig. \ref{fig:fig3}. In particular, the diagram (c) represents the BCS response, while diagram (d) represents the correction due to amplitude or phase fluctuations. Recently, an exact numerical calculations of all these processes in the presence of disorder has been presented for the single-band case\cite{seibold_prb2021}. The relative weight of the two is mainly determined by the level of disorder of the band, expressed as a function of the adimensional ratio $\gamma/(2\Delta)$ of the quasiparticle scattering rate over the gap amplitude. Since the extension of these results to the multiband case is rather challenging, we will construct here a phenomenological model where the frequency dependence of the various contributions is borrowed from the clean-limit calculation, while the coupling of the various modes to light in estimated numerically by the exact results. Such an approach is motivated by the observation that for the disorder level of MgB\textsubscript{2} the simulations of Ref. \cite{seibold_prb2021} have proven that the overall frequency dependence of the third harmonic current induced by both BCS and Higgs fluctuations has a rather similar structure as in the clean case, the only relevant difference being the overall spectral weight of the response.

Let us start from the BCS contribution. We will retain the structure in frequency of the $\sigma_3\sigma_3$ bubble to model the spectrum of the BCS fluctuations, and we will consider the following phenomenological expression for the paramagnetic contributions activated by disorder
\begin{equation}
\label{eq:pbcs}
K^{BCS}_{para}=\sum_{\alpha}w_{\alpha}^{BCS}\chi_{\alpha}^{\sigma_3\sigma_3},
\end{equation}
where $w_{\alpha}^{BCS}$ is a suitable frequency independent weight depending on the level of disorder and on the dispersion of the band. In principle we could also consider the diamagnetic contribution that is still present in a disordered system, but it has been shown that already at small disorder level ($\gamma/(2\Delta)\approx0.1$) it is overwhelmed by paramagnetic processes \cite{seibold_prb2021}. 
For what concerns the Higgs modes we will consider the analogous of (\ref{eq:pbcs}), i.e.
\begin{equation}
K^{H}_{para}=\sum_{\alpha\beta}w_{\alpha}^{H}w_{\beta}^{H}(M_A^{-1})_{\alpha\beta},
\end{equation}
by introducing again proper Higgs-light couplings $w_{\alpha}^{H}$. Finally, the paramagnetic coupling of the Leggett mode with light can be rewritten in terms of a phenomenological weight $w^{L}$, so that
\begin{equation}
K_{para}^{L}=\frac{w^{L}}{L}.
\end{equation}
The overall non-linear response will then be given by 
\begin{equation}
K=K^{BCS}_{para}+K^{H}_{para}+K_{para}^{L}, 
\end{equation}
with suitable effective couplings $w_{\alpha}^{BCS}, w_{\alpha}^{H}$ and $w^{L}$ obtained by a microscopic calculation with disorder, as detailed in the next Section.

\begin{figure}
\centering
\includegraphics[width=\columnwidth]{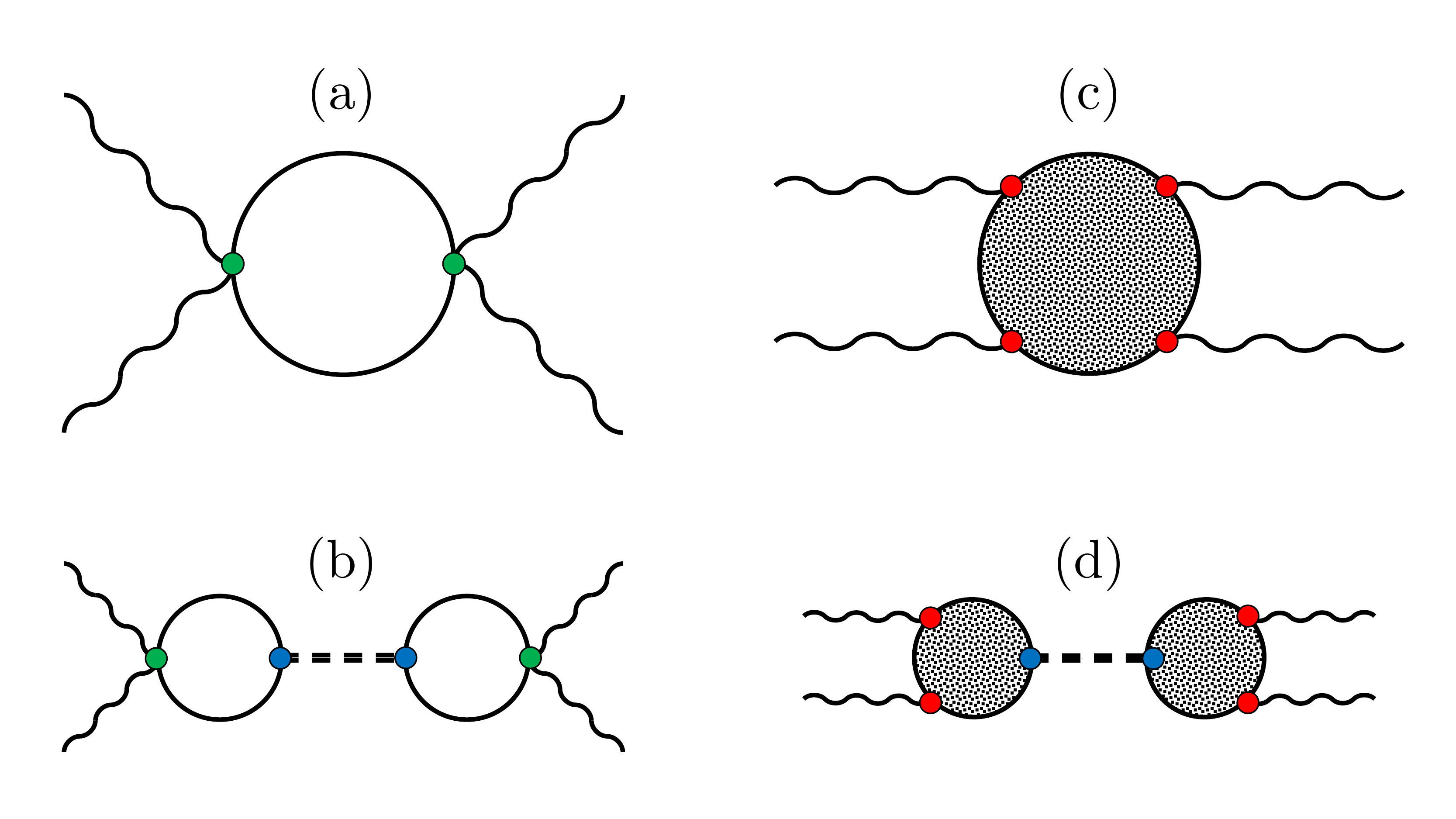}
\caption{The bubbles involved in our phenomenological model of THG in MgB\textsubscript{2}. Wavy lines denote the gauge field $\mathbf{A}$, that does not count in the expressions for $K(\omega)$ reported in the main text, while continuous lines are fermionic Nambu-Green functions in the superconducting state. The green vertices in panels (a) and (b) denote diamagnetic coupling of light with fermions and contain an overall factor $e^2/m_{\alpha}$, while red vertices in panels (c) and (d) denote paramagnetic processes with an overall factor $ev_{\alpha}$. Blurred fermionic bubbles in panels (c) and (d) indicate that in presence of disorder we should take into account the diagram resummation associated with scattering by impurities, as done in \cite{seibold_prb2021}, that is included in our model through the weights $w_{\alpha}$. The blue vertices in panels (b) and (d) represent coupling with collective modes, specifically the Higgs and the Leggett, dashed lines.}
\label{fig:fig3}
\end{figure}

\section{Effective couplings to light in the presence of disorder}
\label{sec:4}

We determine the values of $w_{\alpha}^{BCS}$, $w_{\alpha}^{H}$ and $w^{L}$ by considering the results by Seibold et al. \cite{seibold_prb2021} which treat the effect of disorder \textit{exactly} including every possible vertex and self-energy correction. The starting point is an attractive Hubbard hamiltonian on a square lattice with local disorder
\begin{equation}
H=-\sum_{ij\sigma}t_{ij}c_{i\sigma}^{\dagger}c_{j\sigma}-U\sum_i n_{i\uparrow}n_{i\downarrow}+\sum_{i\sigma}V_in_{i\sigma},
\end{equation}
where $n_{i\sigma}=c_{i\sigma}^{\dagger}c_{i\sigma}$. The first term is the kinetic term with next-neighbor hopping $t_{ij}=t$, the second is an attractive interacting term and the last is a random impurity potential $-V_0\leq V_i\leq V_0$ coupling to the local density, with $V_0$ determining the strength of disorder. The time-evolution of the mean-field density matrix is then computed in presence of a monochromatic vector potential $A(t)$, which allows one to calculate the third order non-linear current $j$ averaging over a certain number of disorder configurations. We refer to the original work \cite{seibold_prb2021} for further details. 

To estimate the effective couplings we compute the non-linear current at zero frequency, considering that its behavior is representative of the disorder dependence of the non-linear response at all frequencies. This quantity is shown in panel (a) of Fig. \ref{fig:fig4}, where we report the results of Ref. \cite{seibold_prb2021} for the BCS response along with its corrections due to amplitude and phase fluctuations, as a function of $\gamma/(2\Delta)$. In Ref. \cite{seibold_prb2021} it was considered a relatively large value of the SC coupling, $U/t=2$. Here, to properly model the case of MgB\textsubscript{2}, we performed the same calculations for a smaller SC coupling $U/t=0.8$ and the disorder level desumed from the experimental conductivity. In general, for a given disorder level, the absolute value of the non-linear current depends both on the band structure and on the value of $\Delta$, being larger for smaller $U$ at weak disorder. Nevertheless, we suggest that the \textit{ratio} of the relative contributions of BCS, Higgs and phase response depends mostly on the disorder level and not on the SC coupling. This can be seen in panel (b) of Fig. \ref{fig:fig4}, where the results at $U/t=0.8$ and $U/t=2$ are compatible within the error bars, once renormalized to have a similar scale. This observation makes it reasonable to assume that the relative strength of the amplitude (phase) couplings with respect to the BCS one $w^{H}/w^{BCS}$ ($w^{P}/w^{BCS}$) are indeed quite universal, and only depend on the disorder level appropriate for each band, $\gamma_{\alpha}/(2\Delta_{\alpha})$. 

\begin{figure}
\centering
\includegraphics[width=\columnwidth]{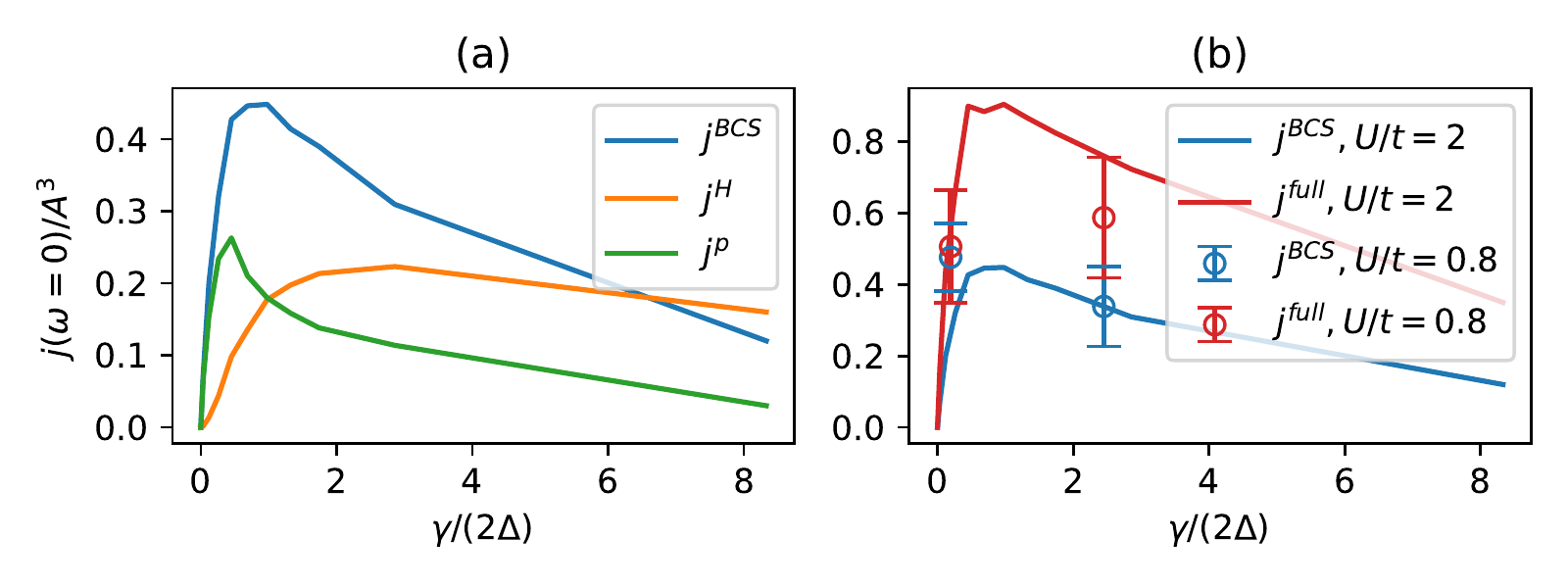}
\caption{In panel (a) we report the non-linear current at zero frequency computed by Seibold et al. in Ref. \cite{seibold_prb2021} for the three separated contributions: BCS (blue line), Higgs (red line) and phase (green line). These results are used to extract the phenomenological weights for the coupling of light to quasiparticles ($w^{BCS}$), to the Higgs modes ($w^H$) and to the phase modes ($w^P$) as a function of disorder. In panel (b) we make a comparison between the zero frequency third order currents allowing only BCS fluctuations (blue line) and all collective modes (red line) in the $U/t=2$ and $U/t=0.8$ cases. The $U/t=0.8$ values are all normalized by the same factor in order to have a similar vertical scale.}
\label{fig:fig4}
\end{figure}

We consider a final correction accounting for the different Fermi velocities $v_{F\alpha}$ of the two bands of MgB\textsubscript{2}, that are of paramount importance in determining the visibility of the contributions from the different order parameters. This can be intuitively justified by realizing that each paramagnetic vertex of the diagrams depicted in Fig. \ref{fig:fig3} carries a factor $ev_{F\alpha}$. This factor is indeed foreseen in analytical approximations at both strong and small disorder levels \cite{silaev_prb2019, haenel_prb2021} for BCS and Higgs fluctuations. For the BCS contribution we then take 
\begin{equation}
w_{\alpha}^{BCS}=e^4v_{F\alpha}^4\tilde{w}^{BCS}\Bigl(\frac{\gamma_{\alpha}}{2\Delta_{\alpha}}\Bigr),
\end{equation}
where $\tilde{w}^{BCS}$ is the value of the zero frequency non linear current reported in Fig. \ref{fig:fig4} at the disorder level $\gamma_{\alpha}/(2\Delta_{\alpha})$. For the Higgs contribution we use instead
\begin{equation}
w_{\alpha}^{H}=e^2v_{F\alpha}^2\zeta_{\alpha}\sqrt{\tilde{w}^{H}\Bigl(\frac{\gamma_{\alpha}}{2\Delta_{\alpha}}\Bigr)},
\end{equation} 
the square root taking into account the presence of two fermionic bubbles in panel (d) of Fig. \ref{fig:fig3}. The $\zeta_{\alpha}$ factor is introduced to account for the softening of the light-Higgs coupling by approaching the critical temperature, and it corresponds to the numerator of (\ref{eq:sig31}), that is indeed proportional to $\Delta_{\alpha}$. We notice that including or not the frequency dependent part of (\ref{eq:sig31}) makes no detectable difference. For the Leggett contribution, mimicking (\ref{eq:kleg}) we take 
\begin{equation}
w^{L}=\Bigl[e^2v_{F\sigma}^2\sqrt{\tilde{w}^{P}\Bigl(\frac{\gamma_{\sigma}}{2\Delta_{\sigma}}\Bigr)}-e^2v_{F\pi}^2\sqrt{\tilde{w}^{P}\Bigl(\frac{\gamma_{\pi}}{2\Delta_{\pi}}\Bigr)}\Bigr]^2\lambda,
\end{equation}
where also here the $\lambda$ factor, being proportional to $\Delta_{\sigma}\Delta_{\pi}$, softens the coupling by approaching the critical temperature. We stress that in the recent analysis of Ref. \cite{shimano_prb19,haenel_prb2021} impurity scattering has been introduced at the level of the Mattis-Bardeen formalism, that corresponds to neglect vertex corrections due to disorder. In this limit the only coupling to the phase is diamagnetic, and then subleading in the presence of disorder. However, as previous work both in the linear\cite{seibold_prb17} and non-linear\cite{seibold_prb2021} regime has shown, the paramagnetic coupling of light to the phase induced by an exact treatment of disorder is crucial at low energy. In the present multiband case our phenomenological model takes into account also this latter type of processes, which are of paramount importance to account for the Leggett-mode contribution to $K$.

\section{Results}
\label{sec:5}

We show in Fig. \ref{fig:fig5} the results of $I_{THG}$ from a monochromatic field at frequency $\omega$ as a function of temperature $T$, corresponding to the modulus squared of the non-linear kernel defined before, $I_{THG}\propto \lvert K(2\omega)\rvert^2$ \cite{cea_prb16}. We focus first on the separate Higgs and BCS contributions, in order to distinguish their effects at the two optical gaps. 

To highlight the importance of the disorder-mediated coupling of the modes to light we also show in panels (a) and (b) the fictious response obtained by using $w_{\alpha}^{H}=w_{\alpha}^{BCS}=1$, i.e. by assuming that the coupling of light is the same for both fluctuation channels in both bands, in order to understand the structure of their spectra in the clean limit. We stress two crucial aspects. First, due to the different absolute values of the order parameters, for a fixed pumping frequency $\omega$ the resonant condition $\omega\approx \Delta_\alpha(T)$ will in most cases occur with only one band, due to the thermal softening of the signal as the temperature increases and the two gap values approach each other. This is due to the presence of the $\sim\tanh(\beta\Delta_{\alpha}/2)$ factor in the numerator of the function $F_{\alpha}(\omega)$, see (\ref{eq:falpha}), that softens the resonance in the region $\Delta_{\alpha}/T\lesssim2$, see dashed line in Fig. \ref{fig:fig0}. Secondly, we notice in panel (b) that at the level of interband coupling of MgB\textsubscript{2}, the spectral weight of amplitude fluctuations is concentrated on the bigger order parameter, with a very faint prominence of the peak in correspondence of the $\pi$ gap. 

In panels (c) and (d) we report the results using the appropriate weights due to disorder. We stress that in order to have a realistic estimate of the ratio $\gamma_\alpha/(2\Delta_\alpha)$ in each band we performed a careful analysis of the absorption in linear response, as detailed in Appendix D, obtaining that $\gamma_{\sigma}=4.5$ meV and $\gamma_{\pi}=6.2$ meV. These values should be contrasted with the values quoted e.g. in Ref. \cite{haenel_prb2021}, where the unrealistic value $\gamma_\pi=100$ meV has been used, that would be far beyond the limit where the concept itself of quasiparticle is valid. For both BCS and Higgs channels the visibility of the $\pi$ gap is enhanced with respect to the $\sigma$ one because the Fermi velocity of the $\pi$ gap is larger, the ratio $v_{F\pi}/v_{F\sigma}\approx1.85$ being extrapolated from ab initio calculations \cite{brinkman_prb2002}. We notice as well that at least at the smaller optical gap the Higgs fluctuations are subdominant with respect to the BCS ones, see the intensity scale for panel (d). In addition, even if disorder enhances the Higgs contribution in the lower $\pi$ band, as we noticed before the Higgs fluctuations of the $\pi$ order parameter are mostly located at the larger optical gap, see panel (e) of Fig. \ref{fig:fig2} and panel (b) of Fig. \ref{fig:fig5}. On the other hand BCS fluctuations in the $\pi$ bands have a maximum at $2\Delta_\pi$, so that once disorder triggers the BCS paramagnetic response in the lower $\pi$ band one recovers a larger signal at the lower optical gap, see panel (c). 

\begin{figure}
	\centering
	\includegraphics[width=\columnwidth]{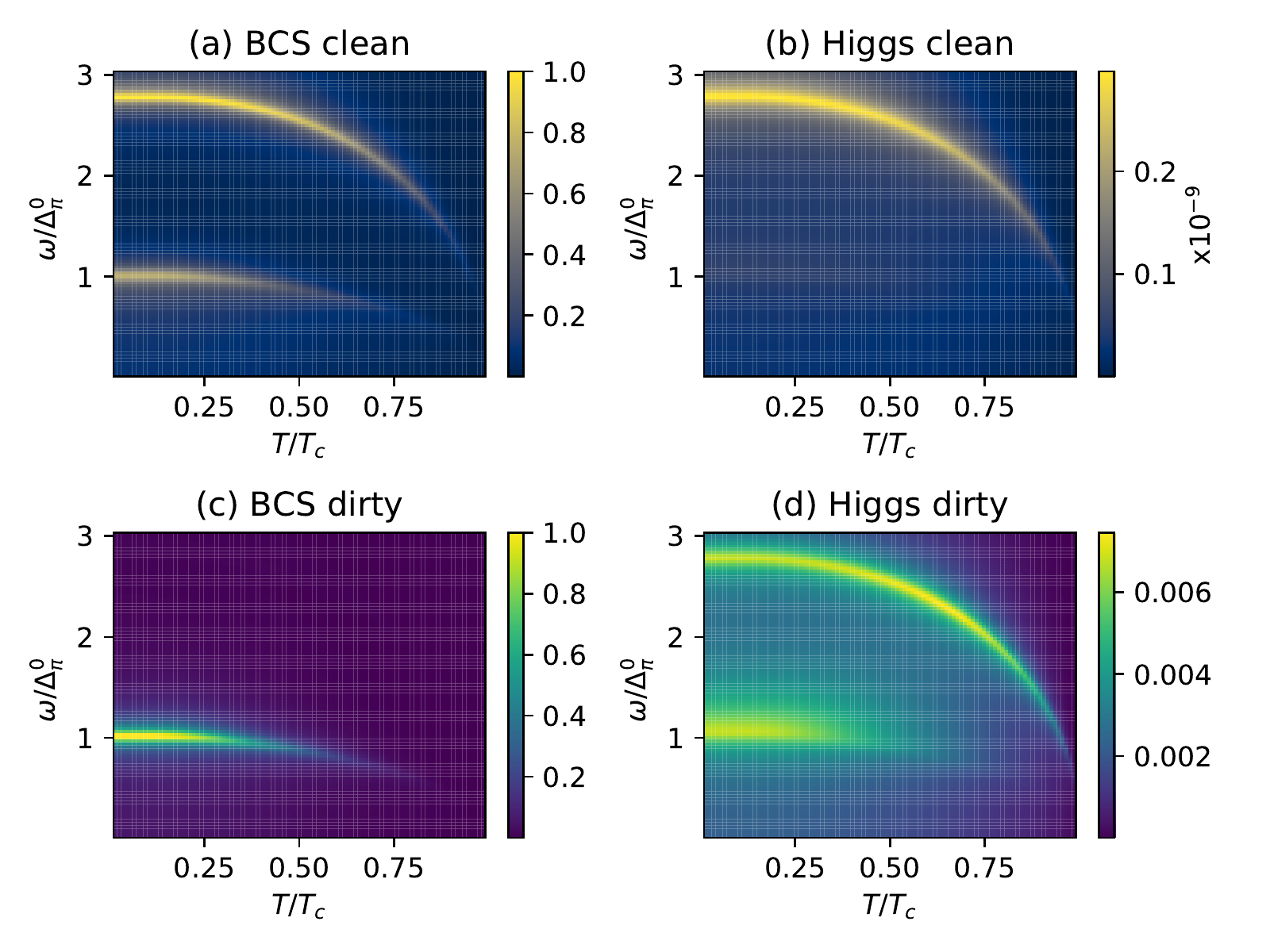}
	\caption{Comparison between the BCS (a) (c) and Higgs (b) (d) contributions to the THG signal as a function of temperature and pump frequency. In panels (a) and (b) the clean diamagnetic response is reported assuming equal band mass, so that the relative visibility of the two bands in the Higgs channel is only determined by the level of interband coupling, see Fig. \ref{fig:fig1}. The prefactor $\Delta/\epsilon_0$ in panel (b) is set to $10^{-3}$. In panels (c) and (d) the dominant paramagnetic response activated by disorder is reported. The stronger visibility of the fluctuations in $\pi$ band is determined by its stronger disorder and larger Fermi velocity.}
	\label{fig:fig5}
\end{figure}

We then performed simulations of the THG intensity including all the contributions of BCS, Higgs and Leggett fluctuations, whose results are shown in Fig. \ref{fig:fig6}. We selected two pump frequencies in order to capture the composition of the THG signal in the two experimental situations of Giorgianni et al. \cite{giorgianni_natphys2019} and Kovalev et al. \cite{kovalev_prb2021}. In the first case we are in resonance with the Leggett mode at $T=0$, where the contribution of the relative phase fluctuations is maximized, see the red horizontal cut in panel (a). As expected, the leading contribution for $\Omega_{P}=1.25$ THz is by far given by the Leggett excitation, while the relatively small bump observed at $T\approx0.8 T_c$ is given by the resonance with the larger $\sigma$ order parameter, as shown in panel (b). Interestingly, at that point the weight of Higgs and BCS fluctuations is comparable. In the second case, the pump frequency $\Omega_{P}=0.5$ THz corresponds to the blue horizontal cut shown in panel (c). In this situation the resonance condition for THG occurs with the smaller $\pi$ order parameter at temperatures where the corresponding BCS part of the non-linear kernel still has a sizable spectral weight, while the resonance with the larger $\sigma$ order parameter occurs for temperatures $T\approx T_c$, where the non-linear kernel is already small due to the thermal effect discussed before. In addition, at this pumping frequency the Higgs contribution is very small since the spectral weight of amplitude fluctuations in the $\pi$ band move to the largest optical gap, as explained before, being far from resonance condition in the experiments of Ref. \cite{kovalev_prb2021}. Also the Leggett gives a vanishingly small contribution, that is not surprising: if the pumping frequency is much lower than half the $T\approx 0$ value of the Leggett mode, the resonance condition $\omega\approx \omega_L(T)/2$ can only occur for $T\approx T_c$, where the Leggett-mode contribution to the non-linear optical kernel is negligible. This is also the condition studied in previous theoretical work\cite{shimano_prb19}, whose conclusions about the irrelevance of the Leggett are only due to the frequency mismatch between the pumping frequency and the Leggett mode and to the lack of inclusion of paramagnetic coupling of light to the phase. In contrast, in the experimental situation of Ref. \cite{giorgianni_natphys2019} the set up has been designed ad hoc to match the condition $2\Omega_P\approx \omega_L(T\approx 0)$ where the Leggett-mode contribution is maximized.

\begin{figure}
	\centering
	\includegraphics[width=\columnwidth]{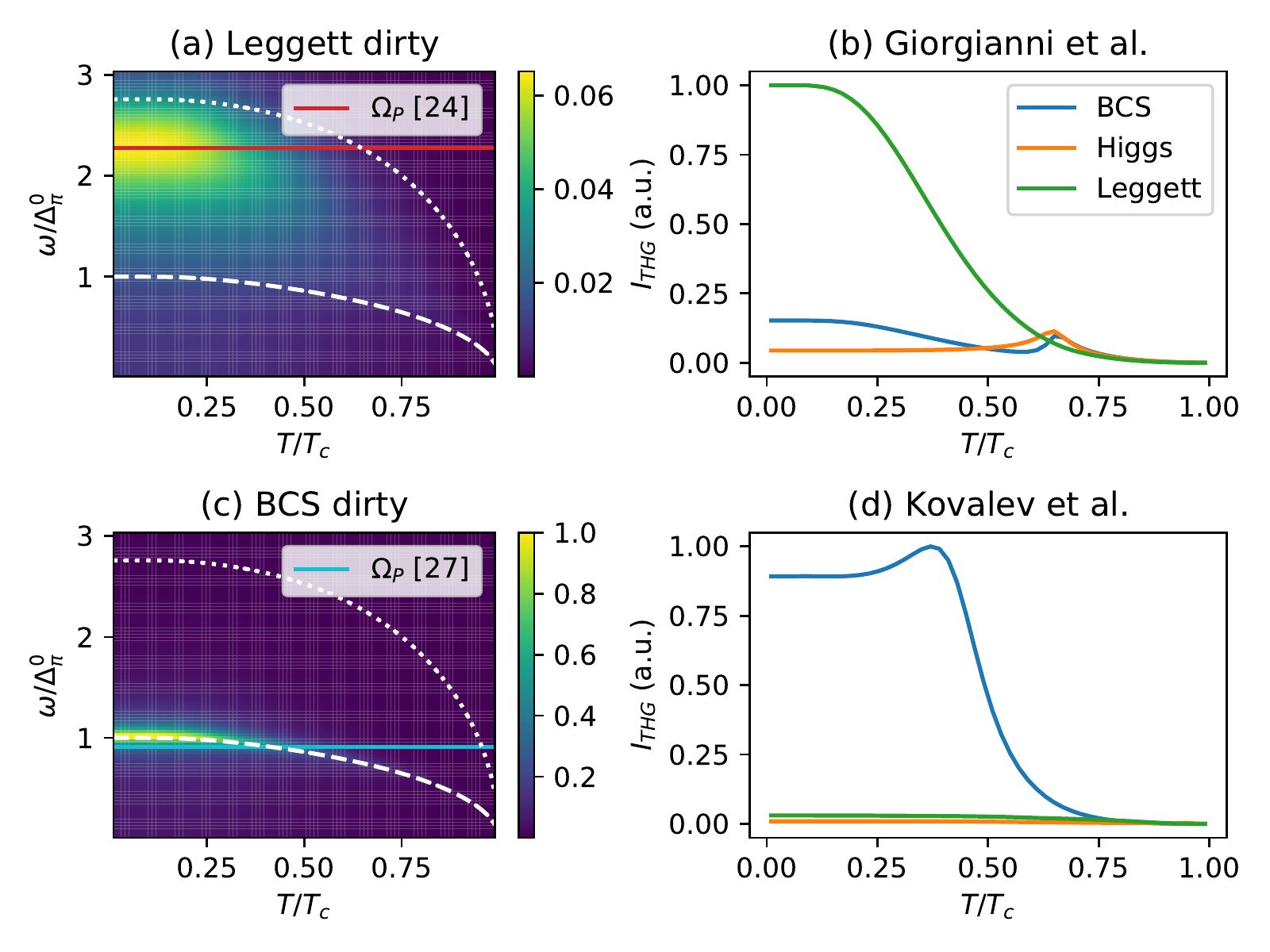}
	\caption{Contribution to the THG signal coming from BCS (blue), Higgs (orange) and Leggett (green) fluctuations for pump frequencies $\Omega_{P}=1.25$ THz (b) used by Giorgianni et al. \cite{giorgianni_natphys2019} and $\Omega_{P}=0.5$ THz (c) used by Kovalev et al. \cite{kovalev_prb2021}. Even thought the $I_{THG}$ has been computed by summing all contributions, in panels (a) and (c) we show the THG response coming from the dominant one alone, along with the temperature dependence of the gaps (white dashed lines) and the corresponding pumping frequency (horizontal solid lines). }
	\label{fig:fig6}
\end{figure}

At last, to make a closer comparison with the experimental results of Ref. \cite{kovalev_prb2021} we account for two additional effects. First, we observe that in order to reproduce the experimental data, reported in panel (d) of Fig. \ref{fig:fig7}, the internal screening of the applied field must be considered. This issue has been recently discussed in Ref. \cite{chu_natcomm2020} within the context of THG measurements in cuprate superconductors. In these systems the $d$-wave symmetry of the order parameter makes the sub-gap optical absorption finite even at very low temperature, leading to a significant temperature dependence of the internal screening. This fact affects the comparison between the experimentally measured THG and the theoretical computation of the non-linear current, that in principle accounts for the response to the internal field, i.e. $j^{NL}(3\Omega_{P})\sim K(2\Omega_{P})\bar{A}_{int}^3$. To account for the difference between the external field, whose spectral components one can easily simulate, and the internal one, that accounts for the response of the material, the authors of Ref. \cite{chu_natcomm2020} suggested to normalize the measured THG to the third power of the measured transmitted first harmonics. For the moment being we use the same approximation and we compute $j^{NL}(3\Omega_{P})\sim K(2\Omega_{P})\bar{A}_{ext}^3t^3$, where $t(\Omega_{P})=\lvert E_{t}(\Omega_{P})\rvert/\lvert E_{ext}(\Omega_{P})\rvert$ is the temperature dependent transmission ratio measured on the same MgB\textsubscript{2} sample in Ref. \cite{kovalev_prb2021}. 

We show in panel (b) of Fig. \ref{fig:fig7} the effect of the internal screening for the $I_{THG}^{BCS}$ from the $\pi$ band obtained for a monocromatic incident field at frequencies $\omega_1=0.3$ THz and $\omega_3=0.5$ THz. For the smaller pump frequency, when the screening effect is neglected (grey dotted line), we observe a small peak at the temperature $T^{*}\approx 0.8 T_c$ such that $\omega_1=\Delta_{\pi}(T^{*})$, followed by an \textit{increase} of the THG signal at lower temperatures. This is again due to the smallness of the $\sim\tanh{(\beta^{*}\Delta_{\pi}/2)}$ factor at the resonance and is ultimately linked to the two-bands nature of MgB\textsubscript{2}, for which $\beta_c\Delta_{\pi}(T=0)\approx0.73$ is much smaller than the single band value $\beta_c\Delta(T=0)\approx1.76$, for which this effect is usually not observed. To reconcile this behavior with the experimental curve, it is then essential to consider the action of screening, the transmission of the material decreasing rapidly near $T_c$ and then being roughly constant down to zero temperature, see grey full line also in same panel. At the larger pump frequency, where correspondingly $T^{*}\ll T_c$, the effect of the screening does not influence the shape of the peak but just contributes to renormalize its overall intensity. 

Finally, we compute the THG intensity by simulating also a realistic narrowband multicycle pulse with form $A(t)=A_0e^{-(t/\tau)^2}\cos{(\Omega_{P}t)}$, where the parameter $\tau$, akin to the temporal duration, is fitted to match the temporal duration of the experimental pulses used by Kovalev et al. \cite{kovalev_prb2021}. The results obtained by including all the fluctuations channel and the internal screening is reported in panel (c) of Fig. \ref{fig:fig7}. We observe an excellent agreement for the grey and blue curves ($\omega_1$ and $\omega_3$ frequencies) while the red one ($\omega_2$) presents an experimentally larger peak than what we found, due to the fact that for $\omega_2$ the resonance condition occurs in a range of temperatures where the largest relative deviations occur between the theoretical gap values and the experimental ones, see panel (a).

\begin{figure}
	\centering
	\includegraphics[width=\columnwidth]{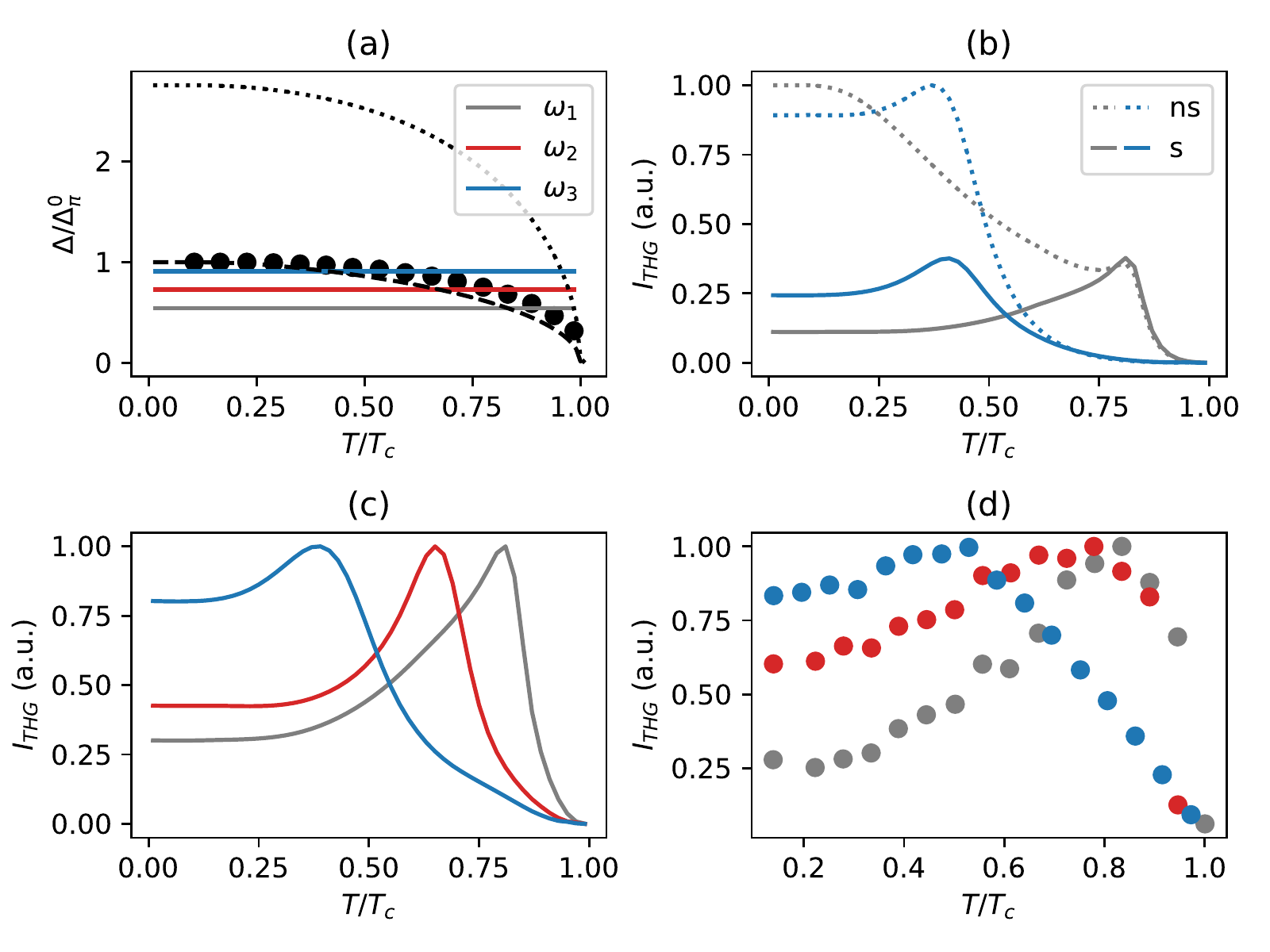}
	\caption{(a) The theoretical results for the temperature dependence of the $\pi$ (dashed line) and $\sigma$ (dotted line) gaps are compared to the experimental data (full circles) obtained by Kovalev et al. \cite{kovalev_prb2021}. The experimental pump frequencies $\omega_1=0.3$ THz (grey line), $\omega_2=0.4$ THz (red line) and $\omega_3=0.5$ THz (blue line) are also reported. (b) BCS contribution to the THG signal from the $\pi$ band for a monocromatic field at the two pump frequencies $\omega_1$ and $\omega_3$ taking into account screening of the applied field (full lines) and neglecting it (dotted lines). (c) Simulated temperature dependence of the THG signal including all fluctuation channels and a realistic pump field. (d) Experimental results of Ref. \cite{kovalev_prb2021} for the same pumping frequencies. The curves at each frequency are normalized to the corresponding maximum THG signal in temperature for a direct comparison.}
	\label{fig:fig7}
\end{figure}

\section{Discussion and conclusions}
\label{sec:6}

In the present manuscript we investigated in a qualitative and quantitative way the non-linear optical response in multiband superconductors, showing that THz spectroscopy represents an excellent tool to investigate the collective modes in this category of superconductors. A first issue, addressed in Section II, concerns the nature of the Higgs modes in a multiband superconductor. In general, while in a single-band superconductor both the quasiparticle BCS continuum and the Higgs mode have a sharp resonance at the optical gap, i.e. at twice the value of the SC order parameter, in the multiband case one can have multiple amplitude fluctuations with different intensity at the two optical gaps. In general, the amplitude fluctuations in a given band $\alpha$ do not necessarily have their maximum at the optical gap $2\Delta_\alpha$ of the same band, due to a subtle interplay between the gap anisotropy and the interband pairing strength. For the specific case of MgB\textsubscript{2} amplitude fluctuations in both bands have larger spectral weight at $2\Delta_\sigma$, i.e. at the larger optical gap, see panel (e) of Fig. \ref{fig:fig1}. This result reflects also on the non-linear optical kernel, where a crucial ingredient is the correct estimate of the coupling of collective excitations to light, in particular in the presence of disorder where paramagnetic-like processes should be considered, as demonstrated by extensive work in the context of single-band superconductors\cite{silaev_prb2019,tsuji_prr20,seibold_prb2021}. Indeed, despite the fact that the lower band is in the relatively large disorder limit, due to the large estimated ratio $\gamma_\pi/(2\Delta_\pi)$ for the samples used in Ref. \cite{kovalev_prb2021}, the Higgs fluctuations of the $\pi$ band have a large spectral weight at $2\Delta_\sigma$, giving overall a small contribution of the Higgs at $2\Delta_\pi$, see panel (d) of Fig. \ref{fig:fig5}. On the other hand, the relatively large $\gamma_\pi/(2\Delta_\pi)$ ratio strongly enhances BCS fluctuations in the $\pi$ band, whose spectral weight piles up at $2\Delta_\pi$ in full analogy with the single-band case. These results fully account for the experimental results of Ref. \cite{kovalev_prb2021}, where the system is driven with three pumping frequencies matching the resonant condition in correspondence of the lower gap $\Delta_\pi$, see panels (c) and (d) of Fig. \ref{fig:fig7}. On the other hand, under this driving conditions the Higgs and Leggett mode give a negligible contribution to the non-linear response.

For what concerns the results of Ref. \cite{giorgianni_natphys2019}, in that case the experiment has been designed with a driving frequency matching half the value of the Leggett mode around zero temperature, namely in the temperature range where the Leggett resonance (that is usually rather broad since it is located above the lower optical gap) is reasonably well defined. In this range of driving frequencies the largest contribution to the non-linear response is due to the Leggett mode, see panels (a) and (b) of Fig. \ref{fig:fig6}, confirming then the original interpretation provided in Ref. \cite{giorgianni_natphys2019}. However, it should be emphasized that in contrast to Ref. \cite{giorgianni_natphys2019} the theoretical estimate of the phase-fluctuations contribution to the non-linear kernel has been based here on the finite paramagnetic coupling of the light to the phase mode, triggered by disorder. This coupling emerges in the exact treatment of disorder implemented in Ref. \cite{seibold_prb2021}, and confirmed here by calculations at smaller SC coupling, while it is absent in the Mattis-Bardeen like treatment of disorder considered in previous theoretical work\cite{shimano_prb19,haenel_prb2021}. Including only the diamagnetic coupling to the phase, as done in Ref. \cite{shimano_prb19}, surely underestimates the Leggett-mode contribution to the non-linear optical kernel. In addition, if one analyzes the THG for a pumping frequency matching the largest gap value, $\omega\approx \Delta_\sigma(T\approx 0)$, as done in Ref. \cite{shimano_prb19}, it is not surprising that the Leggett mode gives a small contribution. This is the direct counterpart on the band $\sigma$ of the results shown in panel (d) of Fig. \ref{fig:fig6} on the band $\pi$, where the Leggett-mode contribution is largely subdominant. Nonetheless, we stress that the overall scale of the THG due to the Leggett mode appears smaller than the one due to the BCS excitations, as one can see by comparing panels (a) and (c) of Fig. \ref{fig:fig6}. To test this prediction experimentally it would be interesting to perform THG measurements on the same sample for a wider range of pumping frequencies, in order to measure the absolute variation of the $I_{THG}$ as different collective modes are excited by the light pulse.

Finally, envisioning possible applications of this analysis, we would like to comment on recent experiments in electron-doped\cite{wang_natcomm21} or hole-doped\cite{grasset_npjqm22} 122 multiband iron-based superconductors. So far, the non-linear response in these systems has been probed via time-domain protocols either with multidimensional THz pulses\cite{wang_natcomm21} or via THz pump-optical probe response\cite{grasset_npjqm22}. In the former case, the authors observed marked oscillations of the transmitted signal as a function of the pump-probe delay $t_{pp}$ at a frequency corresponding to twice the smaller gap. This result\cite{udina_prb19,udina_faraday22} implies a marked resonance of the non-linear kernel at the lowest optical gap, as also suggested by THG experiments in FeSe$_{0.5}$Te$_{0.5}$\cite{shimano_commphys21}. However, to really establish the predominance of the Higgs over the BCS response one would need again to estimate the role of disorder for each band, that was not explicitly provided in Ref. \cite{shimano_commphys21}. Disorder effects were neglected in the theoretical analysis of Ref. \cite{wang_natcomm21}, but the effects of the applied electromagnetic field were included within a non-perturbative scheme, 
so they could access a regime not captured by the theoretical scheme implemented in the present work. The relative importance of the two effects is an interesting open question that deserves further investigation. Finally, in Ref. \cite{grasset_npjqm22} the authors measured the time-resolved signal for a pumping frequency much smaller than both optical gaps, so that one probes in time the square of the applied electric field weighted with $K(\omega\approx 0)$, as theoretically expected by computing the convolution of the kernel with the pump profile\cite{shimano_prl18,udina_prb19,udina_faraday22}. The results confirm the increase of the non-linear response in the SC state, with the additional observation of a larger $B_{1g}$ signal when superconductivity coexists with nematic order. Also in this case, the claim of a relevance of the Higgs is not supported by realistic theoretical calculations: indeed, the authors of Ref. \cite{grasset_npjqm22} do not consider the role of disorder, that is crucial not only to enhance {\em all} the paramagnetic processes, including the BCS ones, but also to change the {\em symmetry} of the non-linear kernel, that is the main focus of Ref. \cite{grasset_npjqm22}. As a consequence, one cannot expect, as claimed in Ref. \cite{grasset_npjqm22}, that the clean-limit calculation reproduces the same polarization dependence of the dirty-limit results. More specifically, detailed recent calculations in Ref. \cite{udina_faraday22} have shown, for the case of cuprates, that the BCS contribution due to paramagnetic processes activated by disorder has a rather different polarization dependence of the diamagnetic one, and both should be considered to properly account for the measured polarization dependence of the signal. We finally observe that the calculation itself of the Higgs-mode contribution presented in Ref. \cite{grasset_npjqm22} has been done assuming only intraband pairing, that is in striking contrast with the predominance of interband pairing in these systems\cite{kotliar_review22,mazin_prl08,chubukov_rev12}. As we explained in Sec. II, to correctly account for the amplitude-mode spectrum in a multiband superconductor it is crucial to use a realistic pairing matrix, and in particular the interband pairing strength has a crucial role in mixing the amplitude fluctuations in the two bands. In addition, in the presence of disorder nematicity has a peculiar effect in mixing the fully symmetric A$_{1g}$ and the B$_{1g}$ channel already in the non-superconducting state\cite{udina_prl20}, as evidenced by measurements of the spontaneous Raman spectroscopy\cite{blumberg_pnas21}. At present, a striking outcome of Ref. \cite{grasset_npjqm22} is that the case of pnictides evidences how THz light pulses can access non-linear optical processes complementary to the ones visible in ordinary Raman measurements. This is probably the most interesting perspective, which motivates future work aimed at establishing the selection rules enconding how light can access the different interaction mechanisms at play in such complex systems. 

\acknowledgments
We acknowledge useful discussions with C. Castellani and J. Lorenzana. This work has been supported by PRIN 2017 No.
2017Z8TS5B, and by Sapienza University via Grant No. RM11916B56802AFE and RM120172A8CC7CC7.

\appendix

\section{Study of the normal state conductivity of MgB\textsubscript{2}.}
\label{app:a}

In order to determine the contribution of the various superconducting modes to the observed THG signal, it is necessary to assess the disorder level in the $\pi$ and $\sigma$ bands of the MgB\textsubscript{2} 13 nm thin film analyzed by Kovalev et al. \cite{kovalev_prb2021}. Here we focus mainly on the available experimental data for the normal state conductivity.

The impurity scattering rates $\gamma_{\sigma}$ and $\gamma_{\pi}$ of MgB\textsubscript{2} have been widely debated in literature. A survey of selected results obtained by modeling linear response functions with one or two Drude terms, plus Lorentz oscillators where included, is reported in Table \ref{tab:survey}. We considered both transmission protocols \cite{jin_apl2005, jung_prb2002, kaindl_prl2001} of THz light through various films and reflectivity measurements \cite{dicastro_prb2006, guritanu_prb2006, hwang_sr2017, kakeshita_prl2006, kuzmenko_ssc2001, tu_prl2001} in a wide spectral range on single crystals, polycrystals and films. While transmission protocols are limited to a few THz region, in the thin film approximation they give immediate access to the complex conductivity without using Kramers-Kronig relations. Conversely, broad spectrum reflectivity protocols require to use the latter to derive the real and imaginary part of linear response functions, but can be combined with ellipsometry, as done in \cite{guritanu_prb2006, kuzmenko_ssc2001}, to obtain more reliable extrapolations even at low frequency. Thus they can both provide a reasonable benchmark of the impurity scattering rates. Other experimental techniques used to infer the impurity scattering rates are Raman spectroscopy \cite{quilty_prl2003} and magnetoresistance \cite{pellecchi_prb2005, pellecchi_prb2006, yang_prl2008}, which usually indicate smaller scattering rates than those coming from linear response functions.

\begin{table*}
\begin{tabular}{cccccccc}
Ref. & Type & Exp. & $\omega_{p,\sigma}$ (eV) & $\omega_{p,\pi}$ (eV) & $\gamma_{\sigma}$ (meV) & $\gamma_{\pi}$ (meV) & Fit \\
\hline
\cite{dicastro_prb2006} & SC & R (300 K) & 2.6 & 4.2 & 37 & 87 & DD+L \\
\cite{guritanu_prb2006} & SC & R (300 K) & $4.14^{*}$ & 4.72 & 12.4 & 85.6 & DD+L \\
\cite{hwang_sr2017} & 1000 nm F & R (50 K) & 2.54 & 1.27 & 6 & 248 & DD \\
\cite{jin_apl2005} & 100 nm F & T (45 K) & - & 4.10 & - & 12 & SD \\
\cite{jung_prb2002} & 50 nm F & T (40 K) & $3.74^{+0.44}_{-0.24}$ & - & $99^{+25}_{-13}$ & - & SD** \\
\cite{kakeshita_prl2006} & SC & R (300 K) & $4.14^{*}$ & $5.89^{*}$ & 25 & 248 & DD+L \\
\cite{kaindl_prl2001} & 100 nm F & T (40 K) & 1.5 & - & 37 & - & SD** \\
\cite{kuzmenko_ssc2001} & PC & R (300 K) & 1.39 & 4.94 & 30 & 1160 & DD+L \\
\cite{tu_prl2001} & 400 nm F & T (45 K) & $1.69\pm0.02$ & - & $9\pm1$ & - & SD** \\
\end{tabular}
\caption{Survey of scattering rates listed in literature. The type labels are single crystal (SC), film (F) and polycrystal (P); the experiment labels are reflectance (R) and transmission (T) with temperature of acquisition of the spectra in brackets; the fit labels are single Drude (SD), double Drude (DD) and Lorentz (L) when present. Values marked with * have been fixed by the authors of the experiment to theoretical ab initio values from \cite{mazin_prl2002}. Single Drude fittings marked with ** do not attribute the observed spectral weight specifically to the $\sigma$ or $\pi$ band, thus the results are arbitrarily listed in the $\sigma$ column.}
\label{tab:survey}
\end{table*}

The spectral range analyzed by Kovalev et al. \cite{kovalev_prb2021} is restricted between 0.5 THz and 2.5 THz, where we expect Drude-like behavior with a constant scattering rate to be dominant. Indeed, the large values of scattering rates obtained in most of the works listed in Table \ref{tab:survey} arise from the attempt to use a constant $\gamma$ in a range of frequency where inelastic effects are relevant. Since here we need to estimate only elastic impurity effects, it is imperative to restrict the analysis to a small range of frequency, as the one provided by the equlibrium measurements of Ref. \cite{kovalev_prb2021}.  Due to the presence of two electronic bands crossing the Fermi surface, in principle we expect both of them to contribute to the normal state conductivity. 

A first indication on the disorder level of the film can then be extrapolated directly from the experimental data. We can interpolate the frequency $\omega_0=1.34$ THz at which the real ($\sigma_1$) and imaginary ($\sigma_2$) part of the conductivity intersect and the corresponding value of the real conductivity $\sigma_1(\omega_0)=3.2\times10^5\,\Omega^{-1}\text{cm}^{-1}$. In the case of a single Drude component $\omega_0$ corresponds to the scattering rate $\gamma$ and the zero frequency conductivity is $\sigma_1(0)=2\sigma_1(\omega_0)$. In our case of two Drude components we can show that irrespectively on the specific scattering rates and plasma frequencies of the two bands we have $\sigma_1(0)>2\sigma_1(\omega_0)$, thus the zero frequency conductivity of this film is \textit{at least} $\sigma_1(0)=6.4\times10^5\,\Omega^{-1}\text{cm}^{-1}$. This value is higher than other previously reported extrapolations from linear response functions \cite{hwang_sr2017, jin_apl2005, kakeshita_prl2006}, thus indicating an extremely clean film. 

It is now imperative to distinguish the $\pi$ and $\sigma$ contributions to the conductivity. We evaluate two different scenarios according to the ratio $\alpha=\gamma_{\sigma}/\gamma_{\pi}$. In pristine MgB\textsubscript{2} samples both theoretical \cite{mazin_prl2002} and all the previously cited experimental studies support the hypothesis that the scattering rate in the $\sigma$ band is smaller than the one in the $\pi$ band, thus we can assume $\alpha<1$. 

In the first scenario we consider $\gamma_{\sigma}\ll\gamma_{\pi}$, as implicitly assumed by Kovalev et al. \cite{kovalev_prb2021} and reported also by Hwang et al. \cite{hwang_sr2017} and Kuz'menko et al. \cite{kuzmenko_ssc2001}, the behavior in frequency is thus completely determined by the $\sigma$ carriers as in a single band case. If we perform a fit of the experimental conductivity with a single band Drude model, the obtained best fit value $\omega_{p,\sigma}=5.1$ eV is notably higher than the theoretical in-plane plasma frequency of the $\sigma$ band $\omega_{p,\sigma}=4.14$ eV \cite{mazin_prl2002}. We stress that including an overall static dielectric constant, as done in \cite{dicastro_prb2006, guritanu_prb2006, kakeshita_prl2006, kuzmenko_ssc2001}, will renormalize the plasma frequency to a still higher value, incompatible with the theoretical prediction. 

To reinforce the rebuttal of this hypothesis, we report a previous study by Zhang et al. \cite{zhang_jap2013} on epitaxial MgB\textsubscript{2} thin films of thicknesses between 6 nm and 40 nm grown by the same hybrid physical-chemical vapor deposition (HPCVD) technique used by Kovalev et al.. The study concluded that the conductivity of those samples is mainly determined by the $\pi$ band plasma. This result has been established by analyzing the behavior of DC resistivity with temperature of the various films and is compatible with the clean sample scenario described with ab initio simulations \cite{mazin_prl2002}. The observed low resistivity at low temperatures and its mild growth with temperature suggest indeed that the scattering rates of $\sigma$ and $\pi$ bands are comparable, thus ruling out the $\gamma_{\sigma}\ll\gamma_{\pi}$ scenario. 

In the second scenario we then consider $\gamma_{\sigma}$ and $\gamma_{\pi}$ comparable. We contemporarily fit the real and imaginary part of the conductivity with the sum of two Drude peaks with different plasma frequencies and impurity scattering rates. Due to the very narrow frequency range accessed by the experiment, we decide to fix the ratio $\omega_{p,\pi}/\omega_{p,\sigma}=1.42$ according to ab initio calculations \cite{mazin_prl2002} and the ratio $\alpha$ of the scattering rates. We fix the value $\omega_0=1.34$ THz at which $\sigma_1$ and $\sigma_2$ intersect and we plot the expected position of the maximum of the imaginary part as a function of $\alpha$ in the inset of Fig. \ref{fig:fig8}. We notice that for $\alpha\lesssim0.5$ the predicted position of the maximum is incompatible with the experimental observation, that occurs in close proximity to or even for a frequency larger than $\omega_0$. We then fix $\alpha=N_{\pi}/N_{\sigma}$ following the prescription of the Born approximation when we assume similar intraband scattering matrix elements of the impurity potential between $\sigma$ and $\pi$ Bloch states. The result of the fitting procedure is shown in the main panel of Fig. \ref{fig:fig8}. We obtain the impurity scattering rates $\gamma_{\sigma}=4.5$ meV and $\gamma_{\pi}=6.2$ meV and the plasma frequencies $\omega_{p,\sigma}=3$ eV and $\omega_{p,\pi}=4.3$ eV.

We notice that both values of scattering rates are smaller than what is reported in Table \ref{tab:survey}, with the notable exception of the result by Jin et al. \cite{jin_apl2005}, that was indeed obtained by a fit in a similar frequency range. As we mentioned before, this is however consistent with the expectation that elastic effects only dominate at low frequencies, so that pursuing a Drude fit with a constant scattering rate in a range of frequencies where inelastic effects dominate leads to an unphysical enhancement of $\gamma$. For what concerns the plasma frequencies, we obtain a smaller value than the theoretically predicted one. The values can be reconciled by introducing a dielectric constant as recalled before or by using an extended Drude formalism with frequency dependent scattering rates \cite{guritanu_prb2006}. Inclusion of inelastic effects leads in general to a larger scattering rate with a consequent increase of the spectral weight. This compensates the small underestimate of the plasma frequencies that we have obtained from such a low-frequency fit.

\begin{figure}
\centering
\includegraphics[width=\columnwidth]{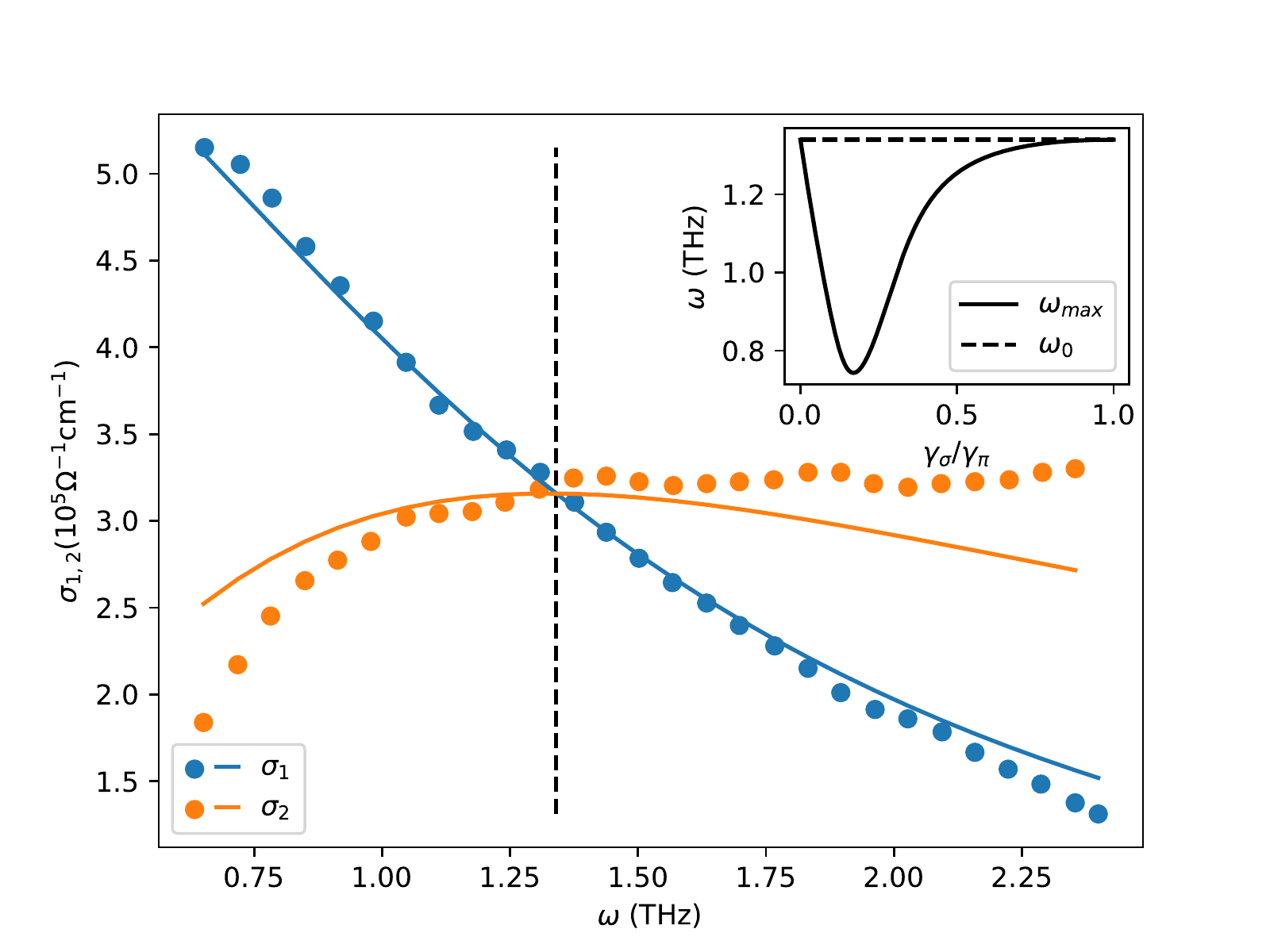}
\caption{Main panel: the fit (full lines) of the real (blue) and imaginary (red) part of the experimental conductivity (full circles) from Kovalev et al. \cite{kovalev_prb2021}. Inset: the expected position of the maximum $\omega_{max}$ of the imaginary part of the conductivity as a function of $\gamma_{\sigma}/\gamma_{\pi}$ (full line) with respect to intersection $\omega_0$ between the real and imaginary part of the conductivity (dashed line).}
\label{fig:fig8}
\end{figure}

%Results from a 100 nm sample attributed the dominant spectral weight in the THz region to the $\pi$ band contribution (CITE JIN), whereas in the present experiment it is attributed to the $\sigma$ band. Previous discussion (CITE ZHANG) about resistivity vs temperature behavior on several film thicknesses suggested to attribute the dominant contribution to the static conductivity to $\pi$ bands when the resistivity is below 5 $\mu\Omega$cm (clean limit) and to $\sigma$ bands when it is over 25 $\mu\Omega$cm (dirty limit). This is consistent with previous first-principle discussion (CITE MAZIN) on the role of intraband and interband impurity scattering in electronic transport. A first extrapolation of the zero-frequency resistivity of the present experimental data ($\rho\sim$ 1.5 $\mu\Omega$cm) indeed suggests a dominant contribution of $\pi$ bands according to the classification just exposed.

\section{Normalizations}
\label{app:b}

There is a certain degree of arbitrariness in defining the weights of our phenomenological model of a two-bands disordered system. To clarify the procedure that we have adopted we start from the single-band description. The paramagnetic BCS contribution at $\omega=0$ and $T=0$ reads 
\begin{equation}
K^{BCS}_{para}=w^{BCS}\chi^{\sigma_3\sigma_3}(\omega,T=0)\equiv\tilde{w}^{BCS},
\end{equation}
since at $\omega=0$ and $T=0$ we want the result to coincide with the non-linear current calculated in \cite{seibold_prb2021}, namely $\tilde{w}^{BCS}$ in the main text. This means that the weight should be normalized as
\begin{equation}
\label{eq:nsb}
w^{BCS}=\frac{\tilde{w}^{BCS}}{\chi^{\sigma_3\sigma_3}(\omega,T=0)}.
\end{equation}
In presence of two bands we want to generalize the previous result. Then considering again that at $\omega=0$ and $T=0$ we have
\begin{equation}
K^{BCS}_{para}=\sum_{\alpha}w_{\alpha}^{BCS}\chi_{\alpha}^{\sigma_3\sigma_3}(\omega,T=0),
\end{equation}
we mimick Eq. (\ref{eq:nsb}) by taking 
\begin{equation}
w_{\alpha}^{BCS}=\frac{e^4v_{F\alpha}^4\tilde{w}_{\alpha}^{BCS}}{\sum_{\beta}\chi_{\beta}^{\sigma_3\sigma_3}(\omega,T=0)},
\end{equation}
that corresponds to take an average weight of the $w_{\alpha}^{BCS}$ by the $\omega,T=0$ values of the $\chi^{\sigma_3\sigma_3}$ clean bubble. A similar procedure is adopted for $w_{\alpha}^{H}$
\begin{equation}
w_{\alpha}^{H}=\frac{e^2v_{F\alpha}^2\zeta_{\alpha}\sqrt{\tilde{w}^{H}}}{\sum_{\beta\gamma}\zeta_{\beta}\zeta_{\gamma}(M_A^{-1})_{\beta\gamma}(\omega,T=0)},
\end{equation}
and $w^{L}$
\begin{equation}
w^{L}=\frac{(e^2v_{F\sigma}^2\sqrt{\tilde{w}^{P}_{\sigma}}-e^2v_{F\pi}^2\sqrt{\tilde{w}^{P}_{\pi}})^2\lambda}{\lambda L(\omega,T=0)}.
\end{equation}

\section{Short-range interaction}
\label{app:c}

Plugging the calculated ab initio coupling matrix into expression (\ref{eq:legmode}), the predicted Leggett-mode frequency is not compatible with both Raman \cite{blumberg_prl07} and pump-probe experiments \cite{giorgianni_natphys2019}. A better agreement is obtained in Ref. \cite{blumberg_prl07} considering an additional repulsive short-range interaction to the hamiltonian
\begin{equation}
H\to H+\sum_{\alpha\beta\mathbf{q}}U_{\alpha\beta}\Phi_{\alpha\mathbf{q}}^{\dagger}\Phi_{\beta\mathbf{q}},
\end{equation}
where $\Phi_{\alpha\mathbf{q}}^{\dagger}$ is written explicitly as
\begin{equation}
\Phi_{\alpha\mathbf{q}}^{\dagger}=\sum_{\mathbf{k}\sigma}c_{\alpha\mathbf{k}+\mathbf{q}/2\sigma}^{\dagger}c_{\alpha\mathbf{k}-\mathbf{q}/2\sigma},
\end{equation}
and $U$ a suitable two by two matrix. It is possible to decouple this interaction in the particle-hole channel by introducing the Hubbard-Stratonovich fields $\rho_{\alpha}$, associated with the density fluctuations of the two bands. An analogous procedure has been adopted in previous work \cite{cea_prb2016} to treat the repulsive Coulomb interaction in the two-bands case. As it has been already pointed out in that context, the phase and density fluctuations are then coupled and cannot be integrated out separately. The action (\ref{eq:genflu}) is then modified as follows
\begin{equation}
S_{\text{fluc}}\sim{\lvert\delta\Delta\rvert}^TM_A\lvert\delta\Delta\rvert+
\begin{pmatrix}
\delta\theta &
\delta\rho
\end{pmatrix}
M_{P,D}
\begin{pmatrix}
\delta\theta\\
\delta\rho
\end{pmatrix},
\end{equation}
having defined the vector $\delta\rho=(\delta\rho_1,\delta\rho_2)$. The matrix of coupled phase and density fluctuations is a straightforward generalization of equation (B8) of Ref. \cite{cea_prb2016} 
\begin{widetext}
\begin{equation}
M_{P,D}=
\begin{pmatrix}
(\frac{i\omega_n}{2})^2\chi_1-\frac{\lambda}{4} & \frac{\lambda}{4} & \frac{i\omega_n}{2}\chi_1 & 0 \\
\frac{\lambda}{4} & (\frac{i\omega_n}{2})^2\chi_2-\frac{\lambda}{4} & 0 & \frac{i\omega_n}{2}\chi_2 \\
-\frac{i\omega_n}{2}\chi_1 & 0 & -\chi_1+2U_{11}^{-1} & 2U_{12}^{-1} \\
0 & -\frac{i\omega_n}{2}\chi_2 & 2U_{12}^{-1} & -\chi_2+2U_{22}^{-1} \\
\end{pmatrix}.
\end{equation}
\end{widetext}
Here we denoted $\chi_{\alpha}=\chi^{\sigma_3\sigma_3}_{\alpha}$. With lengthy calculations it can be shown that the non-zero solution of $\lvert M_{P,D}\rvert=0$ reads
\begin{equation}
\omega^2=\lambda\frac{(\chi_1+\chi_2)}{\chi_1\chi_2}-\lambda\frac{U_{11}+U_{22}-2U_{12}}{2},
\end{equation}
that being $\lambda<0$ in the case of MgB\textsubscript{2} shifts the frequency of the Leggett mode to a higher value, as already stated in Ref. \cite{blumberg_prl07}. There is anyway no physical reason why one should retain $U=g$, i.e. using the same entries for the interaction respectively in the particle-hole and particle-particle channels. In our model we used than the last term of the previous equation as a fitting parameter in order to match the experimental $\omega_L\approx2.5$ THz.

\bibliography{mgb2.bib}

\end{document}